%% file: main.tex
\documentclass[letterpaper,twocolumn,10pt]{article}
\usepackage{usenix-2020-09}
\usepackage{tikz}
\usepackage{amsthm}
\usepackage{amssymb}
\usepackage{amsfonts}

\usepackage{cite}

\usepackage{blindtext}
\usepackage{xcolor}

\usepackage{graphicx}
\usepackage{enumitem}
\usepackage{adjustbox}
\setlist[enumerate]{leftmargin=*}
\setlist[itemize]{leftmargin=*}
\setlist{nolistsep}
\usepackage{amsthm}

\newtheorem{theorem}{Theorem}

\newtheorem{col}{Corollary}
\newtheorem{definition}{Definition}
\newtheorem{prop}{Proposition}
\newtheorem{lma}{Lemma}
\newtheorem{rmk}{Remark}

\usepackage{framed}
\definecolor{shadecolor}{rgb}{0,1,0}

\usepackage[cmex10]{amsmath}
\usepackage{mdframed}

\usepackage{algorithm}
\usepackage{algpseudocode}
\usepackage{algorithmicx}
\usepackage{booktabs}

\usepackage{makecell}
\usepackage{multirow}
\usepackage{enumitem}
\renewcommand{\vec}[1]{\overset{\rightharpoonup}{#1}}
\usepackage{fixltx2e}
\usepackage{lipsum}

\begin{document}
%
\title{Exploring the Security Boundary of Data Reconstruction \\ via Neuron Exclusivity Analysis}

\author{\rm{Xudong Pan, Mi Zhang \thanks{Corresponding Authors: Mi Zhang and Min Yang.}, Yifan Yan, Jiaming Zhu, Min Yang} \\ \textit{School of Computer Science, Fudan University, China} \\ 
\{xdpan18, mi\_zhang, yanyf20, 19210240146, m\_yang\}@fudan.edu.cn 
}

%
%


%



\maketitle

\input{tex/abstract.tex}



%


\input{tex/intro.tex}
\input{tex/related.tex}
\input{tex/prelim.tex}
\input{tex/overview.tex}

\input{tex/atk.tex}
\input{tex/defense.tex}
\input{tex/exp_result.tex}
\input{tex/discussion.tex}

\input{tex/cls.tex}

\bibliographystyle{IEEEtranS}
\bibliography{ref}
%

\input{tex/appendix.tex}

\end{document}

%% file: tex/abstract.tex
\begin{abstract}
Among existing privacy attacks on the gradient of neural networks, \emph{data reconstruction attack}, which reverse engineers the training batch from the gradient, poses a severe threat on the private training data. Despite its empirical success on large architectures and small training batches, unstable reconstruction accuracy is also observed when a smaller architecture or a larger batch is under attack. Due to the weak interpretability of existing learning-based attacks, there is little known on why, when and how data reconstruction attack is feasible.

In our work, we perform the first analytic study on the security boundary of data reconstruction from gradient via a microcosmic view on neural networks with rectified linear units (ReLUs), the most popular activation function in practice. For the first time, we characterize the insecure/secure boundary of data reconstruction attack in terms of the \emph{neuron exclusivity state} of a training batch, indexed by the number of \emph{\textbf{Ex}clusively \textbf{A}ctivated \textbf{N}eurons} (ExANs, i.e., a ReLU activated by only one sample in a batch). Intuitively, we show a training batch with more ExANs are more vulnerable to data reconstruction attack and vice versa. On the one hand, we construct a novel deterministic attack algorithm which substantially outperforms previous attacks for reconstructing training batches lying in the insecure boundary of a neural network. Meanwhile, for training batches lying in the secure boundary, we prove the impossibility of unique reconstruction, based on which an exclusivity reduction strategy is devised to enlarge the secure boundary for mitigation purposes.

\end{abstract}

%% file: tex/intro.tex
\section{Introduction}

From G. Hinton's Turing-award-winning work on \textit{backpropagation} in 1986 \cite{Rumelhart1986LearningRB} to modern optimizers standardized in popular deep learning libraries like Google's Tensorflow \cite{Abadi2016TensorFlowAS} and Facebook's PyTorch\cite{NEURIPS2019_9015}, the \textit{gradient} plays a ubiquitous role in the learning process of most deep learning models. Intuitively, taking the task of image classification for example, the gradient provides the image classifier with a good direction to adapt its parameters for narrowing the errors (i.e., the \textit{loss function}) between the predictions and the ground-truth class labels. As the model iteratively updates its parameters along the opposite direction of the gradient on different training samples, the loss function gradually decreases and the prediction of the learning model becomes more accurate.       

\begin{figure}
\begin{center}
\includegraphics[width=0.5\textwidth]{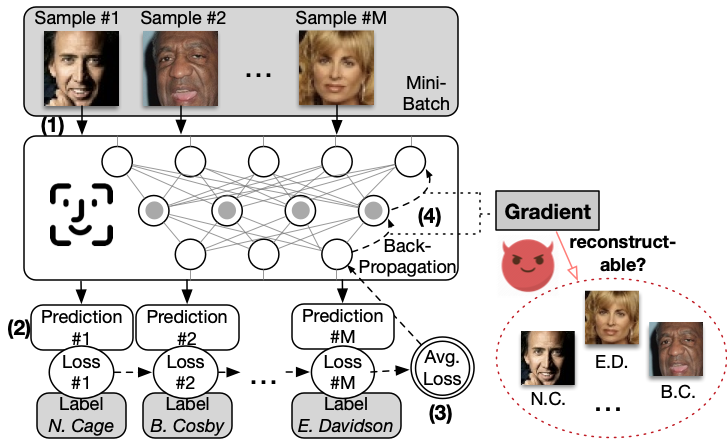}
\caption{The information flow of producing the average gradient of a training batch in a face recognition model and the scenario of data reconstruction attack.}
\label{fig:intro_grad}
\end{center}
\end{figure}

However, accompanied with the fundamental role of gradient in deep learning is its tell-tale heart. As Fig. \ref{fig:intro_grad} shows, in a typical face recognition system, a batch of training images are first input to the neural network classifier. The classifier then predicts the labels, computes the average loss function, and uses back-propagation to calculate the gradient as the parameter derivative of the average loss. As the gradient is explicitly derived from the data inputs and the labels, it is reasonable for an attacker to expect the gradient would leak sensitive information about the original training data. With the booming of novel distributed learning paradigms \cite{yang2019FL,Chen2020NebulaAS}, several research works start to explore the feasibility of inferring the data property \cite{Melis2019ExploitingUF}, the membership \cite{Melis2019ExploitingUF, Nasr2019ComprehensivePA}, the class representatives \cite{Hitaj2017DeepMU, Wang2019BeyondIC}, or the data inputs\cite{Zhu2019DeepLF, Zhao2020iDLGID, Geiping2020InvertingG} from the gradient potentially leaked to a man-in-the-middle attacker or an honest-but-curious server \cite{Lyu2020,Kairouz2021AdvancesAO}.

Despite the feasibility of privacy attacks via the gradient, most previous attacks notice a common yet unclear performance bottleneck on the privacy leakage of their proposed approaches. For example, Melis et al. \cite{Melis2019ExploitingUF} report the precision of sensitive word inference from the gradient decreases by over $30\%$ when the batch size increases by $8\times$. Nasr et al. \cite{Nasr2019ComprehensivePA} report a shallower neural network model is observed to leak less membership information. Zhu et al. \cite{Zhu2019DeepLF} report the iterations required to reverse engineer a training batch from its average gradient increase by $10\times$ when the batch size increases from $1$ to $8$, while the proposed attack is more likely to fail when the neural network is shallow. To summarize, the information leakage from the gradient seemingly decreases for a larger training batch and a shallower neural network model. However, whether this phenomenon has a common root cause interwoven with the underlying mechanism of deep learning? To the best of our knowledge, existing literature provides almost no clue to this fundamental question.


\noindent\textbf{Our Work.} We investigate the above question by dissecting the mechanism of \textit{data reconstruction attack}\cite{Zhu2019DeepLF, Zhao2020iDLGID, Geiping2020InvertingG}, an emerging privacy threat which exploits the leaked average gradient of a deep learning model to reverse engineer the corresponding training batch. As shown in the right part of Fig. \ref{fig:intro_grad}, data reconstruction attack targets at reconstructing the training samples from the corresponding gradient, which poses severe threats on the confidentiality of private training data. As one of the earliest data reconstruction attacks, Zhu et al. \cite{Zhu2019DeepLF} propose a learning-based approach to restore the training batch, which views the unknown training batch as learnable variables (i.e., \textit{dummy data}). By minimizing the L2 distance between the gradient calculated on the dummy data and the ground-truth gradient (i.e., \textit{gradient matching}), they surprisingly observe the reconstruction is possible when the batch size is no larger than $8$ on CIFAR-100 \cite{Krizhevsky2009LearningML}, while the reconstruction quality can be unstable for different trials and relatively small victim models. Follow-up works \cite{Zhao2020iDLGID,Geiping2020InvertingG} present technical adjustments to the learning-based framework in \cite{Zhu2019DeepLF}, with similar bottlenecks observed on data reconstruction. However, due to their weak interpretability, none of the previous works have successfully characterized \textit{why, when and how data reconstruction from gradient is feasible}, which, from our perspective, can be a key entrance to understand and strengthen the privacy properties of the gradient.


To explore the security boundary of data reconstruction from gradient, we present the first analytic study of data reconstruction attacks on the family of fully-connected neural networks (FCNs) with rectified linear units (ReLUs \cite{Goodfellow-et-al-2016}), a quintessential neural network architecture which has been commonly used for demonstrating novel attack and defense insights \cite{Jagielski2020HighAA,Tramr2016StealingML,Rolnick2020ReverseengineeringDR}. As probably the most popular activation function in deep learning practices \cite{Goodfellow-et-al-2016}, a ReLU lets nonnegative inputs pass through without modification and blocks the negative inputs. This special gate-like behavior of ReLU allows each input sample to hold its own set of \textit{activation paths} as its \textit{activation pattern}  \cite{Montfar2014OnTN, Laurent2018TheMS}. We construct deterministic algorithms which decode the hidden information in the average gradient to determine the activation patterns of every single sample, a critical step to reduce the otherwise highly nonlinear gradient-matching problem to a linear equation system regarding the inputs to ease the further analytical studies. Investigating the conditions under which the activation patterns can be reconstructed from the gradient, we mainly make the following key contributions:

\noindent\textbf{(1) Neuron Exclusivity State Analysis.} For the first time, we point out \textit{neuron exclusivity state}, indexed by the number of \textit{\textbf{Ex}clusively \textbf{A}ctivated \textbf{N}eurons} (ExANs, i.e., a ReLU activated by only one sample in a batch during a forward pass), is critical to the feasibility of data reconstruction attack. Specifically, we characterize the following boundary conditions for the neuron exclusivity state of a training batch under attack.

\noindent\textbf{(2) Boundary of Insecure Exclusivity States.} We discover the condition of \textit{sufficient exclusivity}, i.e., when each sample in a batch has at least $2$ ExANs at the last ReLU layer and $1$ at the other layers, as a strong indicator to insecure neuron exclusivity states (Section \ref{sec:atk_case_a}). Specifically, we show a deterministic attack algorithm with guaranteed reconstruction accuracy (Theorem \ref{col:optim_bound_mult}) can be constructed for any training batch satisfying the sufficient exclusivity condition. Evaluation on $5$ real-world scenarios covering medical, face recognition and visual datasets and a diverse set of FCNs of varied depth and width shows, our attack consistently outperforms previous attacks by a large margin in terms of reconstruction recognizability and reaches $100\%$ label inference accuracy (Section \ref{sec:evaluation}). Besides, we also extend our attack algorithm to classifiers based on convolutional neural networks (CNNs) by combining analytical and optimization-based techniques.

\noindent\textbf{(3) Boundary of Secure Exclusivity States.} By dissecting the remaining exclusivity state space, we further determine the \textit{lack of exclusivity} condition, i.e., when each sample has $0$ ExAN at the first ReLU layer, as an indicator to the impossibility of unique reconstruction (Section \ref{sec:def_case_b}). For these states, we prove there always exist infinitely many artifact batches which yield exactly the same gradient as the victim's ground-truth batch, and derive the lower bound for the largest distance between an artifact batch and the ground-truth batch (Theorem \ref{thm:impossible_reconstr}). This observation inspires us to devise an exclusivity reduction strategy, which replaces the first ReLU layer as a linear layer, to enhance the privacy of an arbitrary batch of training samples when its size is larger than the number of neurons in the first layer, with almost no degradation on the model performance. For the completeness of our study, we also present preliminary experimental results in Section \ref{sec:discussion} to empirically analyze the performance of data reconstruction on the remaining states.

%% file: tex/related.tex
\section{Related Work}
\label{sec:related}
\noindent\textbf{Data Reconstruction Attack.} Different from inferring class representatives \cite{Fredrikson2015ModelIA, Hitaj2017DeepMU}, data reconstruction attack primarily aims at recovering each single training sample behind the  intermediate computational results accessed by the attacker. Although  \cite{Salem2019UpdatesLeakDS} first refers to such an attack class as data reconstruction attack, their work mainly study reconstructing a batch of training samples from the changes of their outputs from an updated neural network, which is merely a realistic threat model in most distributed learning paradigms. Parallel to this work, \cite{Wang2019BeyondIC} improves \cite{Hitaj2017DeepMU} with a multi-task GAN to generate individual samples by refining the recovered class representatives, which however requires strong inner-class similarity of the datasets. These limitations make these two attacks not directly applicable to our threat model.  

Recently, starting from \cite{Zhu2019DeepLF}, a branch of research \cite{Zhu2019DeepLF, Zhao2020iDLGID, Geiping2020InvertingG} begins to explore a brute-force yet general approach towards data reconstruction attacks with meaningful empirical results. Solving the gradient matching problem via optimization, these works  mainly differ in the choice of the distance function to minimize (L2 distance in  \cite{Zhu2019DeepLF, Zhao2020iDLGID} and cosine distance in \cite{Geiping2020InvertingG}). Although \cite{Zhao2020iDLGID} uses the property of neural networks to recover the label of a single sample in prior before the learning-based attack, the trick only works for the gradient of a single sample, which makes their method identical to \cite{Zhu2019DeepLF} when applied to the average gradient. Nevertheless, existing attacks mainly stay at an empirical level and aim at showing the feasibility of data reconstruction attacks from the average gradient. Yet, almost no existing works attempt to explain the feasibility and the underlying mechanisms of data reconstruction attack.

\noindent\textbf{Privacy Attacks on Training Data and Beyond.}
As gradients can be more easily accessed in open-network distributed learning systems, a number of recent works begin to study various types of information leakage from gradients \cite{Melis2019ExploitingUF,Nasr2019ComprehensivePA,Hitaj2017DeepMU}. For example,  \cite{Melis2019ExploitingUF} demonstrates the possibility of inferring from the gradient whether the training samples share certain properties (e.g., whether the faces are wtih eye-glasses) and \cite{Hitaj2017DeepMU} leverages a generative adversarial learning paradigm to infer the class representatives, while  \cite{Nasr2019ComprehensivePA} exploits the gradient for membership inference. Different from these existing studies, we are more curious about the feasibility and the theoretical limit of data reconstruction attack, considering its severe threats posed on the private training data \cite{Zhu2019DeepLF}. Besides exploiting the gradient for breaking the training data privacy, researchers also explore, e.g., using the model parameters to infer the properties of training data \cite{Ganju2018PropertyIA, Carlini2019TheSS}, using the intermediate data representations to infer the sensitive attribute values of data samples \cite{Fredrikson2014PrivacyIP,Fredrikson2015ModelIA, Pan2020PrivacyRO}, or using model explanations to reconstruct significant parts of the training set \cite{Shokri2019PrivacyRO}. Aside from training data privacy, previous studies also cover many other aspects of machine learning privacy, including the privacy risks of the data membership \cite{Shokri2017MembershipIA,Salem2019MLLeaksMA,Leino2020StolenML}, the parameters \cite{Tramr2016StealingML}, the hyper-parameters \cite{Wang2018StealingHI}, the model architecture \cite{Duddu2018StealingNN} or its functionality \cite{Orekondy2019KnockoffNS, Jagielski2020HighAA}.  

%% file: tex/prelim.tex
\section{Preliminary}
\noindent\textbf{Gradient in Deep Learning.} Gradient plays an indispensable and ubiquitous role in modern deep learning systems, especially during the model training phase.
In the following, we focus on the $K$-class classification task which covers many real-world use cases of deep learning. We denote a learning model as $f(\cdot; W)$, where $W$ denotes its learnable parameters, and a training sample $(X, Y)$, where $X$ is called the data input and $Y$ is the ground-truth label, ranging in $\{1, \hdots, K\}$. By convention, the learning model takes in the data input $X$ and outputs a vector $f(X;W) \in \mathbb{R}^{K}$ (abbrev. $f$), where the $c$-th element of this vector after a softmax operation predicts the probability of $X$ in class $c$, i.e., $
    p_c := [\text{softmax}(f(X; W))]_{c}= {\exp{f_c}}/{\sum_{c=1}^{K}\exp{f_c}}$, where the operator $[\cdot]_c$ takes the $c$-th entry/row of a vector/matrix, or the $c$-th row of a matrix.

With this prediction, the loss function $\ell(f(X; W), Y)$ (abbrev. $\ell$) is usually calculated as the cross-entropy loss between the predicted probabilities and the ground-truth label, i.e., $\ell(f(X; W), Y) := -\log p_Y = - f_Y + \log{\sum_{c=1}^{K}\exp{f_c}}$. With the aid of modern optimization algorithms (e.g., SGD \cite{Robbins2007ASA} and Adam \cite{Kingma2015AdamAM}), the model parameters are updated along the opposite direction of the gradient, i.e., $\overline{G}(X, Y; W) : = \nabla_{W}{\ell}(f(X; W), Y)$, with a prescribed step size, which guarantees the loss function to decrease iteratively, indicating that the learning model would make more accurate predictions. 

In practice, deep learning systems mainly use the average gradient calculated on multiple training samples (i.e., a batch) for parameter updating, which is usually more suitable for modern parallel computation devices and results in much faster convergence rate \cite{Bubeck2015ConvexOA}. Formally, given a batch of $M$ training samples $\{(X_m, Y_m)\}_{m=1}^{M}$, the average gradient is calculated as the coordinate-wise arithmetic average of the gradients for each single sample, which formally writes  $
        \overline{G}(\{(X_m, Y_m)\}_{m=1}^{M}; W) : = \frac{1}{M}\sum_{m=1}^{M}\nabla_{W}{\ell}(f(X_i; W), Y_i)$.
        


\noindent\textbf{From Gradient Matching to Gradient Equation}. \label{sec:equi_grad_eq}
Existing data reconstruction attacks suppose the attacker captures the average gradient of an unknown batch and has a white-box knowledge about the victim's learning model (i.e., the parameters and the architecture). In practice, such an attacker may be a man-in-the-middle attacker or an honest-but-curious server in distributed learning systems deployed in open networks (e.g., federated learning \cite{Konecn2016FederatedLS}/collaborative training\cite{Chen2020NebulaAS}). Given the leaked average gradient $\overline{G}
$, previous attacks commonly adopt a learning-based approach to solve the following \textit{gradient matching} problem, 
\begin{equation}\label{eq:optim_based}
\min_{\{X_m,Y_m\}_{m=1}^{M}} D(\frac{1}{M}\sum_{m=1}^{M}\frac{\partial\ell(f(X_m;W), Y_m)}{\partial{W}}, \overline{G})
\end{equation}
where $\{X_m, Y_m\}_{m=1}^{M}$ are the learnable variables (i.e., \textit{dummy inputs/labels}) in the gradient matching problem, and a predefined function $D$ measures the distance between the gradient produced by the variables under optimization with the ground-truth average gradient. For example, \cite{Zhu2019DeepLF,Zhao2020iDLGID} implement $D$ as the layerwise L2 distance between the ground-truth gradient and the gradient calculated from the dummy inputs and dummy labels, while \cite{Geiping2020InvertingG} proposes to use the layerwise cosine distance alternatively. Using standard optimizers like L-BFGS \cite{Liu1989OnTL} or Adam \cite{Kingma2015AdamAM} to minimize the learning objective in (\ref{eq:optim_based}) w.r.t. the dummy inputs and labels, one is expected to find a batch of $\{X_m, Y_m\}_{m=1}^{M}$ which yield an average gradient close to the ground-truth gradient. According to the results in \cite{Zhu2019DeepLF, Zhao2020iDLGID, Geiping2020InvertingG}, the authors find the learned dummy inputs are perceptually close to the ground-truth inputs. However, the effectiveness of previous learning-based reconstruction attacks are also observed to  rapidly deteriorate when the batch size $M$ increases and the size of the learning model decreases. 
Yet, there is still little known about the mechanisms which determine this commonly observed yet unclear phenomenon.

In our viewpoint, to optimize the gradient matching problem in (\ref{eq:optim_based}) is equivalent to solve the \textit{gradient equation}:
\begin{equation} \label{eq:grad_eq}
\sum_{m=1}^{M}\frac{\partial\ell(f(X_{m};W), Y_{m})}{\partial{W}} = M\overline{G}
\end{equation}
where $\{(X_m, Y_m)\}_{m=1}^{M}$ are the variables. In other words, the solvability and the uniqueness of the solutions to the gradient equation would largely determine the feasibility of data reconstruction attacks, which is however scarcely explored. 


\noindent\textbf{Fully-Connected Neural Networks with ReLU.} Considering the generality of this open problem, our first analytical study mainly focus on fully connected neural networks (FCNs) with rectified linear units (ReLUs). On the one hand, FCN is a quintessential neural network architecture \cite{Goodfellow-et-al-2016} which is commonly used for demonstrating novel attack and defense insights \cite{Jagielski2020HighAA,Tramr2016StealingML,Rolnick2020ReverseengineeringDR}, and a popular choice for classification tasks on data samples in vector form or feature vectors extracted from upstream feature extraction models \cite{Ji2018ModelReuseAO}. On the other hand, due to its numeric stability \cite{Goodfellow-et-al-2016}, ReLU is commonly implemented in a very broad class of popular neural network architectures including both FCNs and deep convolutional neural networks (CNN). Intuitively, a ReLU $\sigma(\cdot)$ can be viewed as a gate structure which allows non-negative values to pass through without any change and meanwhile blocks negative values by outputting $0$ instead, which is formally written as $
\sigma(x) = x$ if $x\ge{0}$; $\sigma(x)=0$ if $x<{0}$. 

For simplicity, we refer to an FCN with ReLU as an FCN. Formally, an $(H+2)$-layer FCN has the following formulation $
f(X;W_0, W_1,\hdots, W_H, b_0, b_1, \hdots, b_H) = W_H\sigma(W_{H-1}\hdots(W_1\sigma(W_0X+b_0)+b_1)\hdots+b_{H-1}) + b_H$, where $W_i \in \mathbb{R}^{d_{i+1}\times{d_i}}$ is the weight matrix at the $i$-th layer, $b_i \in \mathbb{R}^{d_{i+1}}$ is the bias vector, the data input $X \in \mathbb{R}^{d_0}$, $d_{H+1} = K$, i.e., the class number, and $\sigma$ is the ReLU activation function. For example, when $H=1$, the model $W_1\sigma(W_0X+b_0)+b_1$ is called a three-layer FCN. We denote an FCN architecture in the form of $(d_0$-$d_1$-$\hdots$-$d_{H+1})$. Moreover, without loss of generality, we would omit the bias terms in our analysis for the simplicity of notations. As Appendix \ref{sec:app:config} shows, the bias terms are reduced to constant calibration after the original gradient equation is simplified to a linear equation system. 

\begin{figure}[ht]
\begin{center}
\includegraphics[width=0.5\textwidth]{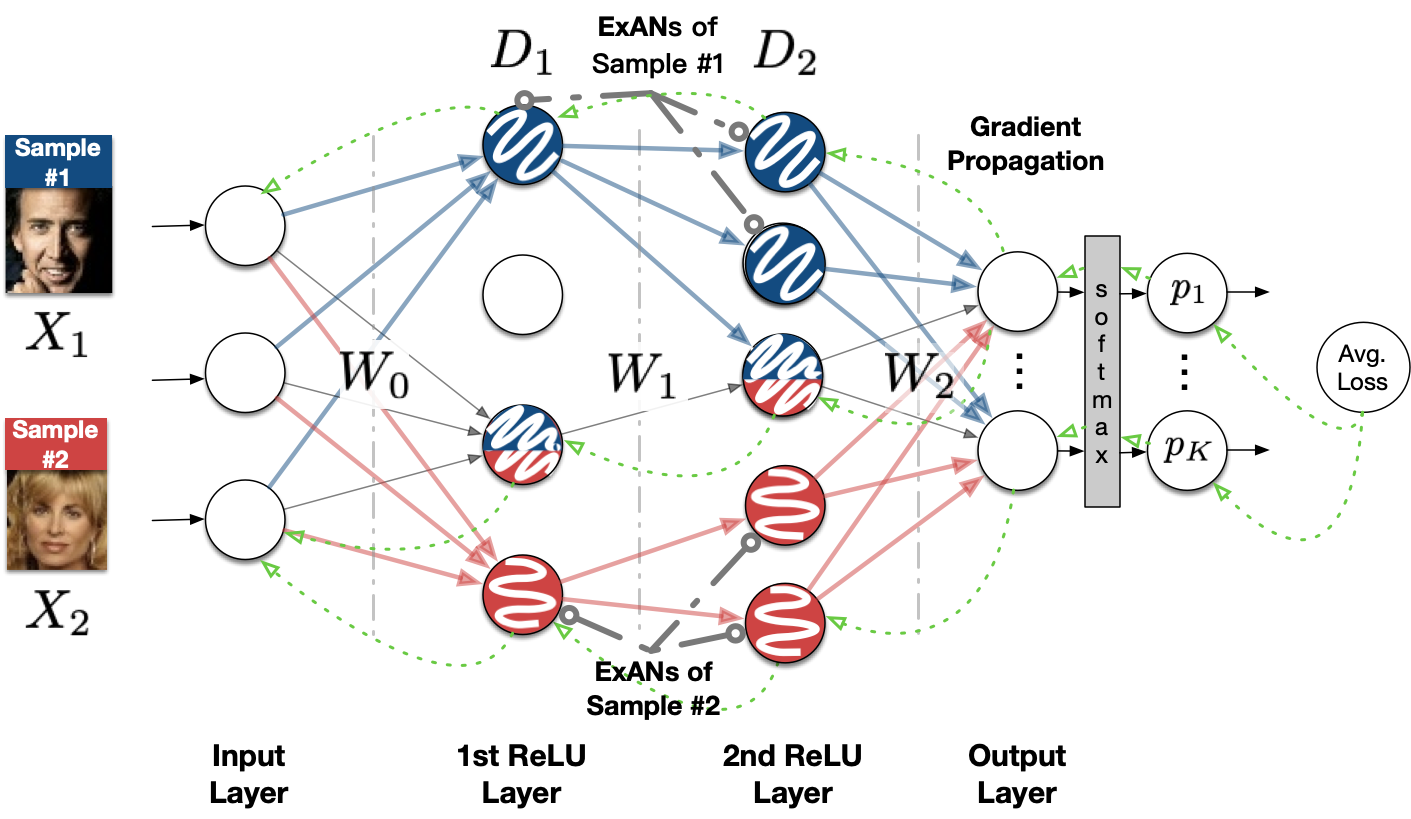}
\caption{The forward and backward phase of a training batch in a $4$-layer FCN (better viewed in color).}
\label{fig:activ_pattern}
\end{center}
\end{figure}

\noindent\textbf{Activation Patterns.}
Considering the gate-like behavior of ReLU, when a representation is input to a neural network with ReLU, each coordinate of the representation selectively passes through a part of neurons at the current layer and meanwhile is blocked by the remaining neurons due to the negativity or a vanishing weight of the neural connection. As Fig. \ref{fig:activ_pattern} shows, after forwarding through the whole neural network layer by layer, each sample has a set of computation paths in the neural network, which forms its \textit{activation pattern}. Below, we develop the idea of activation pattern in a formal way.   


ReLU is applied to a vector in a coordinate-wise way. For example, the $i$-th output of the first layer, i.e., $\sigma(W_0X+b_0)$,  is reformulated as $
\sigma(W_0X+b_0) := D_1(X;W_0, b_0)(W_0X+b_0)$ \cite{Laurent2018TheMS}, where $D_1(X;W_0, b_0) = \text{diag}(\mathbf{1}\{\sigma(W_0X+b_0) \succ 0\})$, i.e., a diagonal matrix whose $j$-th diagonal entry is $1$ when the $j$-th output of the first layer is positive and otherwise $0$. For simplicity, we denote the last term as $D_1(X)(W_0X+b_0)$. We call such a matrix $D_1(X)$ the \textit{activation matrix} of $X$ at the first layer. Similarly, we can reformulate the whole ReLU FCN as $
f(X) = W_HD_{H}(W_{H-1}\hdots(W_1D_1(W_0X+b_0)+b_1)\hdots + b_{H-1})+b_H$
where the sequence of activation matrices $(D_1, \hdots, D_H)$ describes the \textit{activation pattern} for the data input $X$.

Finally, we would like to mention a useful property of the activation pattern during the gradient back-propagation, that is, the activation matrix commutes with the derivative operation, i.e., $\nabla_{W_0}D_i(X)W_{i-1}\hdots W_0X = D_i\nabla_{W_0}W_{i-1}\hdots W_0X$. In other words, the gradient backpropagates along the same activated path as in the forwarding phase. Fig. \ref{fig:activ_pattern} illustrates the role of the activation pattern in the forward and the backward phases of an FCN, where each data sample is forwarded through a set of computation paths which composes its \textit{activation pattern} $(D_1, D_2)$. For example, as the blue directed lines show, Sample $\#1$ passes through the 1st and the 3rd neuron at the first ReLU layer, which means its activation matrix at the first layer $D_1$ is $\text{diag}(1, 0, 1, 0)$. Similarly, at the second ReLU layer, its activation matrix $D_2$ is $\text{diag}(1, 1, 1, 0, 0)$. Moreover, we call a neuron which is only activated by one sample in an input batch as the \textit{exclusively activated} neuron (i.e., ExAN) of the corresponding sample (marked in the same color of the sample). For simplicity, the green dashed lines plot parts of the back-propagation paths: the gradient signal is non-vanishing only along the same activation pattern in the forward phase.



%% file: tex/overview.tex
\section{Overview of Analytic Framework}
\noindent\textbf{Threat Model.}
As summarized in Table \ref{tab:threat_model}, we follow almost the same threat model as in existing data reconstruction attacks \cite{Zhu2019DeepLF, Geiping2020InvertingG}, where the attacker has the knowledge of:
\begin{enumerate}
    \item The ground-truth average gradient $\overline{G}$ calculated on a batch of $M$ training samples. 
    \item The architecture and the parameters of an FCN with respect to which the gradient is calculated.
\end{enumerate}
Unlike previous attacks, we do not require the knowledge of the batch size $M$. As Section \ref{sec:atk_case_a} will show, the attacker can determine the batch size from the gradient alone for certain exclusivity states. In our analysis on the boundary conditions, we additionally assume the model has random weights to ensure an attacker cannot exploit the otherwise trained parameters for better attacks. Nevertheless, we later show this assumption has no influence on the effectiveness of our proposed attack.  




\noindent\textbf{Summary of Key Results.} As one of our major contributions, we for the first time unveil and prove the strong relation between the feasibility of data reconstruction attacks on FCNs and the \textit{exclusivity} of activation patterns of samples in a batch (i.e., \textit{neuron exclusivity state}), indexed by the number of \textit{\textbf{Ex}clusively \textbf{A}ctivated \textbf{N}eurons} (\textit{ExAN}s) in each ReLU layer. First, we formally define what is an ExAN.

\input{tex/tables/threat_model}

\begin{definition}[ExAN]
Given a batch $\{(X_m, Y_m)\}_{m=1}^{M}$, we call the $j$-th neuron at the $i$-th layer is an ExAN if $\sum_{m=1}^{M}[D_i(X_m)]_{j} = 1$, where $[D_i(X_m)]_j$ denotes the $j$-th diagonal entry of the activation pattern of $X_m$ at the $i$-th layer. 
\end{definition}
Literally, an ExAN is a ReLU activated by only one sample in a batch during the forward pass. For intuition, Fig. \ref{fig:activ_pattern} illustrates two data samples and their corresponding ExANs during their computation in a four-layer FCN. We further denote the number of ExANs for the $m$-th sample at the $i$-th layer as $N_i^m$, which is calculated as $n(\{j: [D_i(X_m)]_{j} = 1 \bigwedge \forall{m^{'}\neq {m}}, [D_i(X_m)]_{j} = 0\})$, where $n(\cdot)$ denotes the cardinality of a set.  
Based on the definition, we present the following boundary conditions which provide sufficient conditions for both the insecure and the secure neuron exclusivity states respectively.

\begin{itemize}
    \item \textbf{Insecure Boundary Condition.} \textit{(Sufficient Exclusivity)}: $\mathit{N_H^m \ge 2}$ and $\mathit{\forall{i=1, \hdots, H-1}}$, $N_{i}^m \ge 1$. Intuitively, the condition of \textit{sufficient exclusivity} characterizes that each sample in a batch has at least $2$ ExANs at the last ReLU layer and has at least $1$ ExAN at the other ReLU layers. We call such a batch as an \textit{insecure} batch. In this case, we present in Section \ref{sec:atk_case_a} the construction of a deterministic attack algorithm which has guaranteed reconstruction accuracy and stably outperforms previous attacks in evaluation (Sections \ref{sec:evaluation:attack}). 
    
    \item \textbf{Secure Boundary Condition.} \textit{(Lack of Exclusivity)}: $\mathit{N_{1}^m = 0}$ and $\mathit{M > d_1}$. As a contrast, the condition of \textit{lack of exclusivity} covers the situations when each sample in a batch activates the same set of neurons in the first layer. In this case, we prove the impossibility of unique reconstruction based on the gradient only, and correspondingly derive a simple yet effective privacy enhancing strategy based on a slight modification on the FCN architecture (Section \ref{sec:def_case_b}). 
\end{itemize}


%% file: tex/tables/threat_model.tex
\begin{table*}[ht]
  \centering
  \caption{Summary of threat models of different data reconstruction attacks.}
  \scalebox{0.9}{
    \begin{tabular}{lcccc}
\toprule
          & \textbf{DLG} \cite{Zhu2019DeepLF} & \textbf{iDLG} \cite{Zhao2020iDLGID} & \textbf{Inverting} \cite{Geiping2020InvertingG} & \textbf{Ours} \\
         \midrule
    \textbf{Target Architecture} & {Unspecified} & {Unspecified} & {Unspecified} & {\makecell[l]{FCN/Extensible to CNN}} \\

    \textbf{Attack Technique} & Optimization & Optimization & Optimization & Analytic/Hybrid \\
      {\makecell{\textbf{Type of Leaked Gradient} \\ \textit{(Average/Single-Sample)}}} & Both  & Single-Sample & Both  & Both \\
    \textbf{Batch Size is Required?} & Required &  N/A  & Required & Not Required \\

    \bottomrule
    \end{tabular}}%
  \label{tab:threat_model}%
\end{table*}%

%% file: tex/atk.tex
\section{Reconstruction under Sufficient Exclusivity}\label{sec:atk_case_a}

In this section, we present a novel deterministic algorithm for reconstructing an unknown insecure batch $\{(X_m, Y_m)\}_{m=1}^{M}$ from the average gradient $(\overline{G}_{0}, \hdots, \overline{G}_{H})$ with guaranteed accuracy.

\noindent\textbf{Gradient Equation of an FCN.}
As mentioned in the first part of Section \ref{sec:related}, the loss function $\ell_m := \ell(f(X_m), Y_m)$  is usually implemented as the cross-entropy between the ground-truth label $Y_m$ and the ``softmax-ed'' $f(X_m)$. With simple calculations, the gradient of the entropy loss on the $c$-th output of $f(X_m)$, i.e., $f_c^{m}$, has the following closed form: 

\begin{equation}\label{eq:g_def}
{\partial\ell_m}/{\partial{f_c^{m}}} = \overline{g}_c^{m} = -1+p_c^{m}\text{ if }c=Y_m \text{ else }p_c^{m} \quad{},
\end{equation}
where $p_c^{m} := p_c(X_m)$ is the predicted probability for the sample $X_m$ in class $c$. For convenience, we use the \textit{loss vector} $\overline{g}^{m}$ to denote $ (\overline{g}_1^{m}, ..., \overline{g}_K^{m})$.

Based on the chain rule, the gradient of $W_i$ (i.e., the weight of the $(i+1)$-th layer) contributed by the $m$-th sample is $\nabla_{W_i}{\ell_{m}} = \sum_{c=1}^{K}\overline{g}_c^{m}{\nabla_{W_i}{f_c^{m}}}$. By summing over $m$ and replacing the left side as the captured gradient at the $i$-th layer, i.e., $\overline{G}_i$, we have the following gradient equation for $W_i$, $
M\overline{G}_i = \sum_{m=1}^{M}\sum_{c=1}^{K}\overline{g}_c^{m}\nabla_{W_i}{f_c^{m}}$,
which provides a highly complicated nonlinear equation system for the attacker to solve, where the nonlinearity lies in $\overline{g}_c^{m}$ and the activation patterns $D_i(X_m)$ (or, concisely, $D_i^{m}$) contained in $f_c^{m}$.

\noindent\textbf{Simplification to Linear Equation System.} Under the condition of sufficient exclusivity, we show both $\{(\overline{g}_c^{m})_{c=1}^{K}\}_{m=1}^{M}$ and $\{(D_i^{m})_{i=1}^{H}\}_{m=1}^{M}$ can be uniquely determined to reduce the nonlinear gradient equation above to a linear equation system.

\textbf{(1) Inference of Loss Vectors:} First, to infer $\overline{g}_c^{m}$, we consider the gradient equation for $W_H$, i.e., 
\begin{equation}
M[\overline{G}_H]_c = \sum_{m=1}^{M}\overline{g}_c^{m}f^{m}_{H-1}, 
\end{equation}
where $f^{m}_{H-1} := D_H^{m}W_{H-1}...D_1^{m}W_0X_m$. We discover the following sufficient condition for recovering $\overline{g}_{c}^{m}$.

\begin{prop}
\label{prop:reconstruct_g}
A sufficient condition for determining the ratio of $\overline{g}_c^{m}$ over $\overline{g}_{1}^{m}$ is, each data sample has at least two ExANs at the last but one layer. 
\end{prop}

\begin{figure}[ht]
\begin{center}
\includegraphics[width=0.4\textwidth]{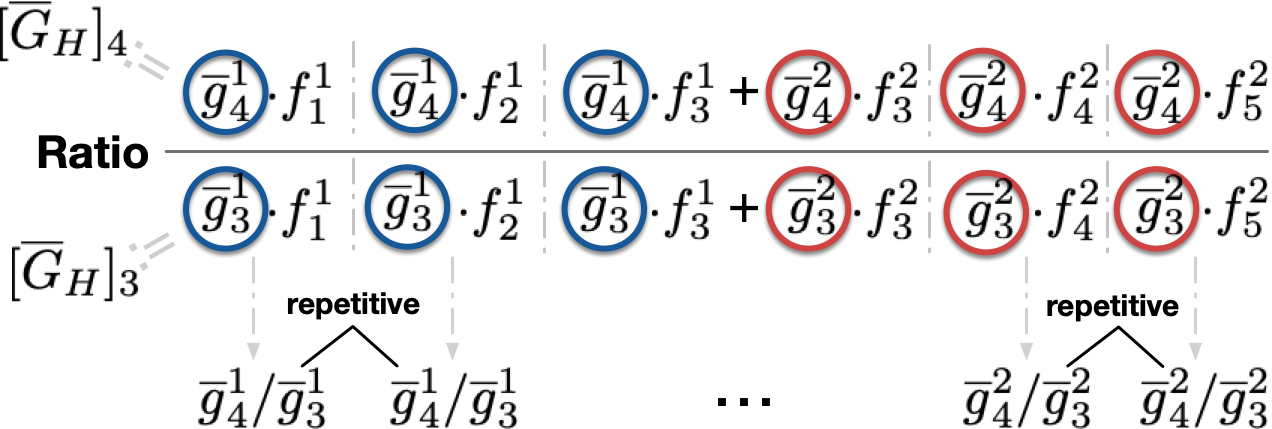}
\caption{A schematic proof on the observation that ExANs at the last ReLU layer help solve the ratios among $\{g_c^{m}\}_{c=1}^{K}$ for each $m$.}
\label{fig:exclu_active}
\end{center}
\end{figure}

As a proof, we construct the following algorithm to determine the ratios $\{(\overline{g}_{c}^{m}/\overline{g}_{1}^{m})_{c=2}^{K}\}_{m=1}^{M}$. For better intuition, we consider the case in Fig. \ref{fig:activ_pattern} where each sample $X_m$ in a batch of size $2$ has two ExANs at the last layer and one commonly activated neuron (i.e., $X_1$ takes up the 1st and the 2nd neurons, and $X_2$ the 4th and 5th). In other words, both samples activate two different neurons at the last ReLU layer of the neural network. According to the gradient equation above, by forming the ratio vector $[\overline{G}_H]_4/[\overline{G}_H]_3$, we notice that for each ExAN of the 1st sample (i.e., the 1st \& the 2nd neuron), the element $[[\overline{G}_H]_4/[\overline{G}_H]_3]_1 = [[\overline{G}_H]_4/[\overline{G}_H]_3]_2 = \overline{g}_4^1/\overline{g}_3^1$. We also provide a schematic proof of this property in Fig. \ref{fig:exclu_active}. 
Based on this property, we can practically detect the repetitive values in $[\overline{G}_H]_c/[\overline{G}_H]_1$ to determine the ExANs for the $m$-th sample and then collect the value at the corresponding index of the ratio vector $[\overline{G}_H]_c/[\overline{G}_H]_1$ as the corresponding ratio $\overline{g}_c^{m}/\overline{g}_1^{m}$. Similarly, by enumerating the class index $c$, we can again reduce the $M\times{K}$ variables in  $\{(\overline{g}_{c}^{m})_{c=1}^{K}\}_{m=1}^{M}$ to $K$ variables. Below, we present two noteworthy remarks on inferring the label and determining the concrete values of $\overline{g}_c^{m}$ based on the ratio equations. For more implementation details, please refer to Algorithm \ref{alg:determine_gcm}.

\begin{rmk}[Exact Label Inference] \label{remark:label_reconstr} From (\ref{eq:g_def}), only if $c$ hits the ground-truth label $Y$, then $g_c$ is negative while the others are positive. This observation is also noticed by \cite{Zhao2020iDLGID} independently. As a result, by checking the signs of the recovered ratios, the attacker can easily determine the ground-truth label $Y$ of the data $X$. For details, please see Algorithm \ref{alg:determine_label}.   
\end{rmk}

\begin{rmk}[Feasible Range of $\overline{g}_1$]
Moreover, with the constraint that $\sum_{c\neq Y}\overline{g}_c = \sum_{c\neq Y}p_c\le 1$, we can determine the feasible range $[0, \delta]$ of $\overline{g}_1$, where $\delta$ is a rather small constant in practice, which allows the attacker to use a random value in the range or run binary search to get satisfying results. Below, it is reasonable to assume $\overline{g_1}$ is known. 
\end{rmk}

\textbf{(2) Inference of Activation Patterns:}
Based on the knowledge of the ExANs at the last ReLU layer, we present the following exclusivity condition under which the attacker can uniquely determine the activation pattern $(D_i^{m})_{i=1}^{H}$ for each data sample. 

\begin{prop}
\label{prop:reconstruct_D}
Given the knowledge on the ExANs at the last ReLU layer, the attacker can determine $\{(D_i^{m})_{i=1}^{H}\}_{m=1}^{M}$ with uniqueness, if each data sample $X_m$ has at least one ExAN in $D_{i}^{m}$, $i \in \{1, ..., H-1\}$.
\end{prop}
Below, we provide a brief algorithmic proof. In general, the procedure of determining the activation patterns is recursively done from the last to the first ReLU layer. Initially, we have already recovered at least two ExANs in $D_H^{m}$ for each input $X_m$. Therefore, if we consider the $j$-th neuron as the ExAN for $X_m$, then the $j$-th column of $\overline{G}_{H-1}$ only consists of the gradient w.r.t. $X_m$. Hence, by checking the non-zero positions of the $j$-th column, we immediately get the diagonal terms of $D_{H-1}^{m}$. Similarly, with the $(H-1)$-th layer solved, the procedure can be done for the $(H-2)$-th layer, and so on, until the first layer. Readers may refer to Fig. \ref{fig:activ_pattern} for better intuition. Meanwhile, the attacker can further determine the whole $D_H^{m}$ for each $m$-th sample by solving the gradient equation w.r.t. the last bias vector $b_{H-1}$ via dynamic programming. Details on the above algorithm can be found in Algorithm \ref{alg:determine_dm}.

\noindent\textbf{An Upper Bound on Reconstruction Errors.}
After the loss vectors and the activation patterns are determined, the nonlinear gradient equation collapses to a system of linear scalar equations, which can be solved with off-the-shelf linear equation solvers (e.g., LSMR \cite{ChinLung2011LSMRAI}). For the implementation details, please refer to Appendix \ref{sec:app:config}. When the gradient equation is reduced to a linear form, the reconstruction error is influenced by the number of scalar linear equations available to the attacker and the number of samples the attacker wants to solve. Specifically, for an attacker who solves the least-square-error solution of the linear gradient equation as an approximation to the victim's ground-truth data inputs, we derive the following error upper bound of data reconstruction.

\begin{theorem}[Reconstruction Error Bound]
\label{col:optim_bound_mult}
 Under the insecure boundary condition, when the sparsity of the gradient at each $i$-th layer satisfies $1-\beta(\overline{G}_i) < \epsilon_i{\frac{\sqrt{d_id_{i+1}}}{M\dim{\mathcal{X}}}}$ (Note: $\beta(\overline{G}_i)$ denotes the ratio of non-zero elements in the full gradient), then the attacker can reconstruct the labels exactly and recover the ground-truth data inputs $\{X_m^{*}\}_{m=1}^{M}$ within the following mean square error bound:
\begin{align}\label{eq:error_bound_mult}
\frac{1}{M}\sum_{m=1}^{M}\|X_m - X_m^{*}\|_2 <
O({\sum_{i=0}^{H}\epsilon_i(1-\beta(\overline{G}_i))}(\sum_{m=1}^{M}\|X_m^{*}\|_2))
\end{align}
\end{theorem} 
Omitted technical proofs are all provided in Appendix \ref{sec:app:proof}. Intuitively, Theorem \ref{col:optim_bound_mult} details the quantitative relation between the upper bound of the average reconstruction error and several key characteristics about the victim. For example, when the gradient information provided to the adversary is sparser, the batch size or the dimension of the problem space is larger, then the $\epsilon_i$ increases according to the inequality in the premise, which in turn makes the error bound at the RHS of (\ref{eq:error_bound_mult}) larger and hence causes the reconstruction quality less stable. On the contrary, when the layer width $d_i, d_{i+1}$ are enlarged and the gradient information stays at a similar level, the $\epsilon_i$ decreases and therefore the attacker can expect a smaller reconstruction error bound. 


\noindent\textbf{Extension to Convolutional Neural Networks.} When attempting to extend the above analytical results to convolutional neural networks (CNNs), we notice \textit{the weight parameters are shared among each input dimension for a convolutional layer but not for a linear layer} would inhibit a direct extension. Although a convolutional layer is mathematically equivalent to a sparse fully-connected layer in the forward phase, the former has a rather different behavior from a sparse linear layer during the backward phase, as the gradient signals conceptually propagated to each dimension of the weight of the equivalent sparse fully-connected layer are actually accumulated to the same weight parameter in the convolutional filter. In this situation, we could neither check the non-zero/zero elements in the gradients to determine the activation state of each neuron in the feature map, nor to determine the ExANs for each data sample, which inhibits the reduction of the otherwise nonlinear gradient equation to a solvable linear equation system. Moreover, even if the reduction were possible, the number of scalar gradient equations provided by convolution filters can be highly insufficient to form a determined equation system with a satisfying solution.  

\begin{figure}[t]
\begin{center}
\includegraphics[width=0.5\textwidth]{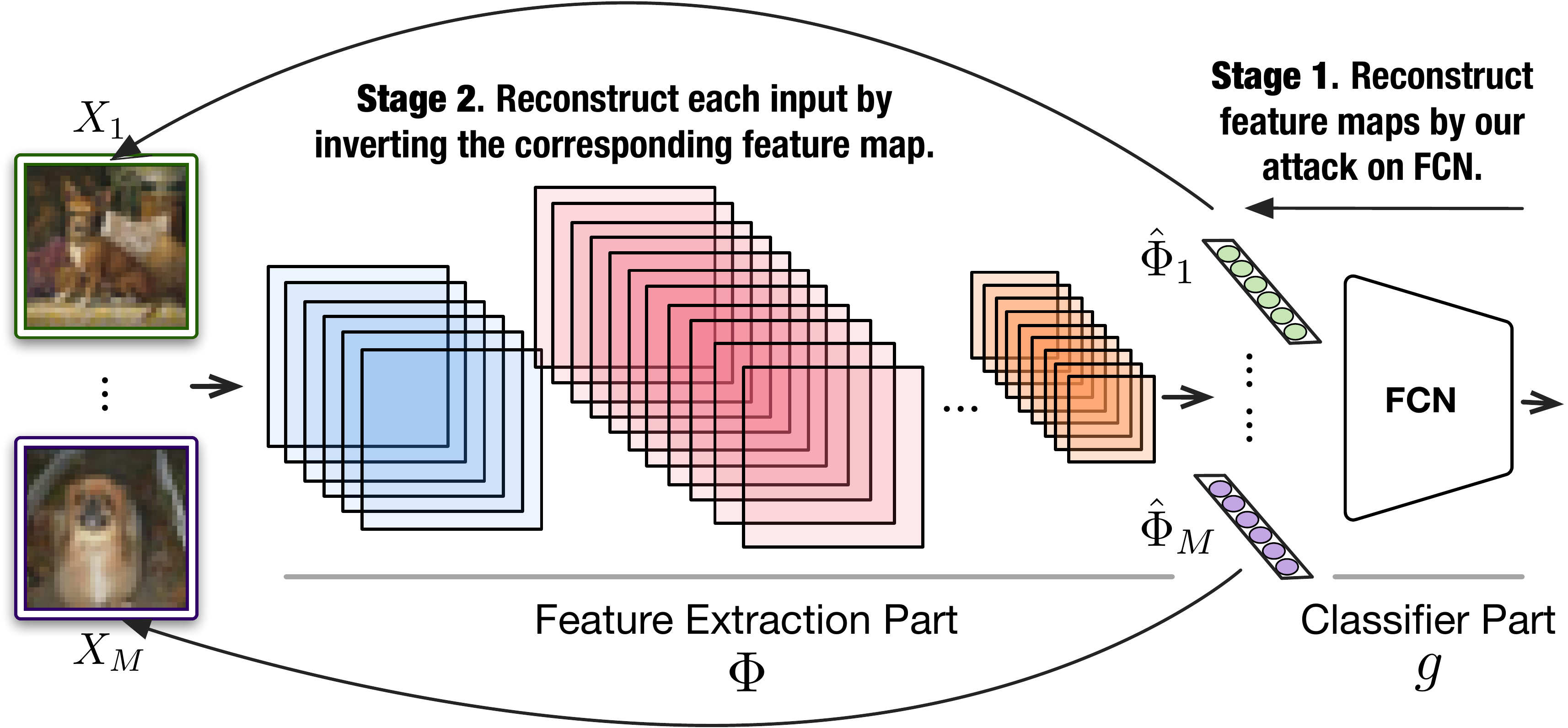}
\caption{Overview of our hybrid attack on CNN-based classification models.}
\label{fig:ac2_attack_pipeline}
\end{center}
\end{figure}

In this work, we alternatively extend our proposed attack algorithm on FCN as a two-stage hybrid approach towards data reconstruction attacks on CNN-based classification models. As a mild assumption, we assume the target CNN-based classification model can be decomposed into explicitly as $f = \Phi\circ{g}$, where $\Phi$ is a feature extraction model mainly composed of convolutional and pooling operations, and $g$ is an FCN for classification. This characterizes a common practice of CNN models in the real world \cite{Simonyan2015VeryDC,Ji2018ModelReuseAO}. As Fig. \ref{fig:ac2_attack_pipeline} shows, our extended attack pipeline contains the following stages:

\noindent$\bullet$\textbf{ Stage 1.} At the first stage, we reconstruct the inputs (i.e., the feature maps) to the FCN $g$. Based on our obtained results on FCNs, the feature maps of each sample can be reconstructed with guaranteed reconstruction accuracy, under the condition of sufficient exclusivity, which we denote as $\{\hat{\Phi}_1, \hdots, \hat{\Phi}_M\}$.

\noindent$\bullet$\textbf{ Stage 2.} At the second stage, with the reconstructed feature maps, we aim to solve the input-output constraint $h(X_m) = \hat{\Phi}_m$ for each $m=1, \hdots, M$ with gradient-based optimization algorithms. This is equivalent to an optimization problem 
$
        \text{arg}\min_{X_m}\|\Phi(X_m) - \hat{\Phi}_m\|^2.
$
   We name the optimization problem as the \textit{feature matching problem} to distinguish it from the \textit{gradient-matching problem} in (\ref{eq:optim_based}) solved in learning-based data reconstruction attacks. In our implementation, we utilize the technique proposed in an interpretability-related work \cite{Ulyanov2018DeepIP}, which mainly models the variable $X_m$ as the output of a trainable neural network $h(\cdot; \psi):\mathbb{R}^{d}\to\mathcal{X}$ on a fixed random noise $z$, corresponding to the following optimization objective of our hybrid attack:
 \begin{equation}
        \text{arg}\min_{\psi}\|\Phi(h(z_m; \psi)) - \hat{\Phi}_m\|^2,
\end{equation}
    where the optimization is conducted on the parameters of the model $h$. For more implementation details, please refer to Appendix \ref{sec:app:config}. 

As a final remark, we highlight the tight relation of our proposed hybrid attack on CNN with our analytic and attack techniques on FCN. On the one hand, our hybrid attack still exploits the key condition of sufficient exclusivity to separate out and reconstruct the feature map of each individual sample. From our perspective, how to separate the information of each single data sample which is otherwise mixed in the average gradient is critical to the feasibility of data reconstruction attacks. On the other hand, without our attack algorithm on FCNs to reconstruct the feature map for each sample to a tolerably small error, one cannot bootstrap the otherwise challenging task of data reconstruction from the average gradient to the feature matching problem, a more simplified task as we further discuss in Appendix \ref{sec:app:config}. 

%% file: tex/defense.tex
\section{Privacy Enhancement via Exclusivity Reduction}\label{sec:def_case_b}
\noindent\textbf{Impossibility Results under Lack of Exclusivity.} First, we show the lack of exclusivity leads to the impossibility of unique reconstruction, i.e., \textit{given a ground-truth batch of data samples $\{X_i\}_{i=1}^{M}$, there always exist an infinite number of artifact batches which have exactly the same gradients as the ground-truth one.} 
\begin{theorem}[Impossibility of Reconstruction]
\label{thm:impossible_reconstr}
For an FCN $f(X) = W_H\sigma(W_{H-1}\hdots(W_1\sigma(W_0X+b_0)+b_1)\hdots+b_{H-1}) + b_H$ s.t. $d_1 < d_0$ and a batch of samples $\{X_i\}_{i=1}^{M}$, if $N_{1}^m = 0$ and $M > d_1$, then there always exists a linear space $\mathcal{Q} \subseteq \mathbb{R}^{{d_0}\times{M}}$, which satisfies: $\forall \Delta \in \mathcal{Q}$ and $\forall{i} = 1, \hdots, H$,
\begin{align}
    \overline{G}(\{(X_m, Y_m)\}_{m=1}^{M}; W_i) = \overline{G}(\{(X_m+\Delta_m, Y_m)\}_{m=1}^{M}; W_i) \label{eq:weight_inv} \\ 
    \overline{G}(\{(X_m, Y_m)\}_{m=1}^{M}; b_i) = \overline{G}(\{(X_m+\Delta_m, Y_m)\}_{m=1}^{M}; b_i) \label{eq:bias_inv}
\end{align}
Moreover, when the input space has the interval constraints $X+\Delta \in [-1, 1]^{d}$ (common for the image domain), the L2 norm of the largest perturbation $\Delta$ has the following lower bound, 
\begin{equation}
    \|\Delta\|_2^2 \ge \sum_{i=1}^{M}\|{\eta}_i\|_2^2 - \text{Tr}(A^{\dagger}AY^{T}Y), 
\end{equation}
where $A = [\alpha_1^T, \dots, {\alpha}_M^T]$, $\vec{\alpha}_m = \sum_{c=1}^{K}\overline{g}_c^{m}([W_H]_c^{T}D_H^m \hdots W_1D_1^m)$, $Y = [{\eta}_1^{T}, \dots, {\eta}_1^{T}]$, and ${\eta}_i = |TP_0X_i|$ (where $|X|$ takes the absolute values of entries in $X$), with $P_0 = (I-W_0^{\dagger}W_0)$ and the columns of $T$ are the left singular vectors of $W_0$. 
\end{theorem}
In other words, Theorem \ref{thm:impossible_reconstr} indicates, without additional information, each artifact batch is indistinguishable from the ground-truth batch for the adversary in our threat model. 
\begin{figure}[t]
\begin{center}
\includegraphics[width=0.45\textwidth]{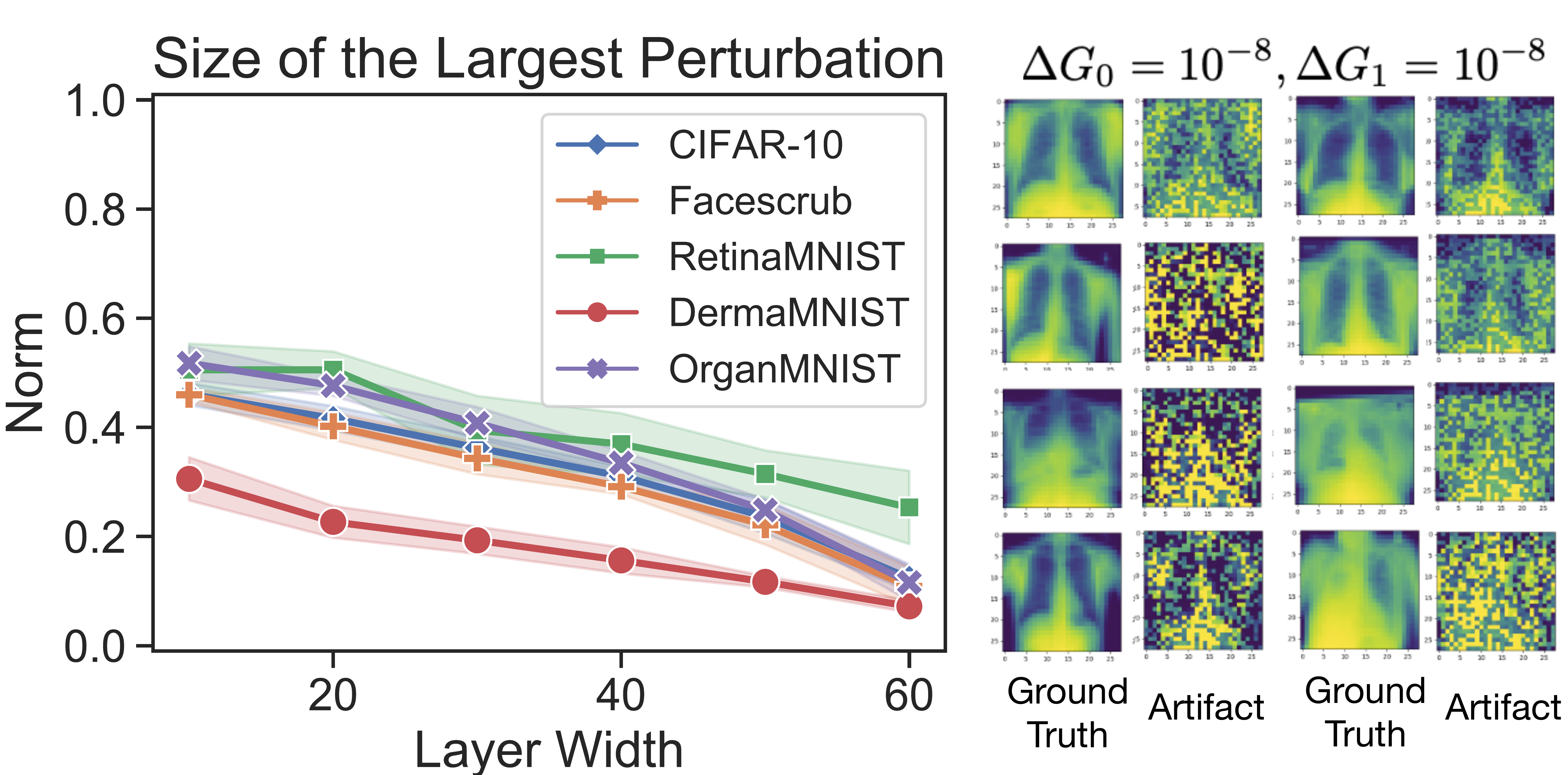}
\caption{\textbf{Left}: The empirical values of the largest perturbation (per dim.) available in the perturbation subspace when the layer width varies, where the range of the y-axis marks the largest possible modification to an input dimension while making it stay in $[-1, 1]$. \textbf{Right}: The artifact batches which share the same gradient (to a $10^{-8}$ numeric error) with the ground-truth batches on OrganMNIST.}
\label{fig:ac5_theory}
\end{center}
\end{figure}
At the left of Fig. \ref{fig:ac5_theory}, we report the empirical values of the lower bound of the largest perturbation norm, where the largest perturbation that can be added to one pixel without changing the gradient is as large as $0.5$ when the width of the first layer is $10$, which forms a $25\%$ relative deviation compared with the $[-1, 1]$ range of a pixel's value. At the right of Fig. \ref{fig:ac5_theory}, we further visualize a batch of size $8$ when the largest perturbation is added to the ground-truth data samples while preserving the average gradient calculated on a $(d_0$-$7$-$512$-$K$) FCN. As is shown, almost each single input can be obfuscated to an unrecognizable level while the average gradient of the obfuscated batch differs from the ground-truth one by an $10^{-8}$ numeric error. Combining the results above, we expect the existence of the perturbation subspace with a considerable size under the lack of exclusivity will have a positive effect on inhibiting the attacker from reconstructing useful information from the average gradient only.


\noindent\textbf{Enhancing Gradient Privacy by Exclusivity Reduction.}
Although the above impossibility result under the lack of exclusivity poses a natural defense against data reconstruction attacks, we however notice with experiments that the situation of a batch of samples sharing the same activation pattern at the first hidden layer rarely happens. To utilize the above observation, we propose the exclusivity reduction strategy below to modify the conventional FCN architecture for ensuring the lack of exclusivity and thus the impossibility of unique reconstruction.
\begin{col}[Exclusivity Reduction]
\label{col:exclusivity_reduction}
When we remove the first ReLU layer in a conventional FCN, i.e.,
\begin{align}
W_H\sigma(W_{H-1}\hdots(W_1\hat{\sigma}(W_0X+b_0)+b_1)\hdots+b_{H-1}) + b_H
\end{align}
where $\hat{}$ denotes the omission of the term, then, for a batch of samples $\{X_i\}_{i=1}^{M}$ s.t. $M > d_1$, there always exists a linear space $\mathcal{Q} \subseteq \mathbb{R}^{M\times{d_0}}$ such that for each $\Delta \in \mathcal{Q}$ and ${i} = 1, \hdots, H$, Theorem \ref{thm:impossible_reconstr} holds.
\end{col}
The motivation behind is straightforward: after the first ReLU layer is removed, every sample in a batch activates all the neurons in the first layer, which naturally guarantees the lack of exclusivity. Consequently, according to Theorem \ref{thm:impossible_reconstr}, we can construct infinitely many artifact batches which are considerably different from the ground-truth batch in perception yet indistinguishable in terms of the gradients (Fig. \ref{fig:ac5_theory}). The details on constructing the artifact batch can be found in the proof of Corollary \ref{col:exclusivity_reduction} in Appendix \ref{sec:app:proof}. Further, we show in Fig. \ref{fig:acc_plot} that such a modification would cause almost no performance degradation for the practical usage of FCNs. 

As a final remark, exclusivity reduction is essentially different from collapsing the first two layers (e.g., $W_0 \in \mathbb{R}^{d_1\times{d_0}}, W_1 \in \mathbb{R}^{d_2\times{d_1}}$) into a single layer (i.e., $\tilde{W_1} = W_2, \tilde{W_0} = W_1W_0\in\mathbb{R}^{d_2\times{d_0}}$), because, in the backward phase, the gradient information accessible to the attacker becomes $\nabla_{W_1W_0}{{\ell}(X, Y)}$ after exclusivity reduction, which provides at most $d_0\times{d_2}$ scalar equations to solve, instead of  $\nabla_{W_0}{\ell}(X, Y)$, $\nabla_{W_1}{\ell}(X, Y)$, which brings at most $(d_0 + d_2) \times{d_1}$ equations to solve.


%% file: tex/exp_result.tex
\section{Evaluation Results}
\label{sec:evaluation}
\input{tex/exp_setting.tex}

\subsection{Attacks inside Insecure Boundary} 
\label{sec:evaluation:attack}
\noindent\textbf{Comparison of Reconstruction Accuracy.} We compare the performance of our proposed data reconstruction attack with two previous attacks, i.e., DLG \cite{Zhu2019DeepLF} and Inverting \cite{Geiping2020InvertingG}, on each scenario in Table \ref{tab:scenarios}, where the target FCN architecture is ($d$-$512$-$K$) and the batch size $M=8$. We do not involve iDLG \cite{Zhao2020iDLGID} as it is only applicable to gradient calculated on a single sample (Table \ref{tab:threat_model}). The \textbf{FCN} rows in Table \ref{tab:exp_full_table} compare the performance of our proposed attack with the baselines.

\begin{figure}[ht]
\begin{center}
\includegraphics[width=0.45\textwidth]{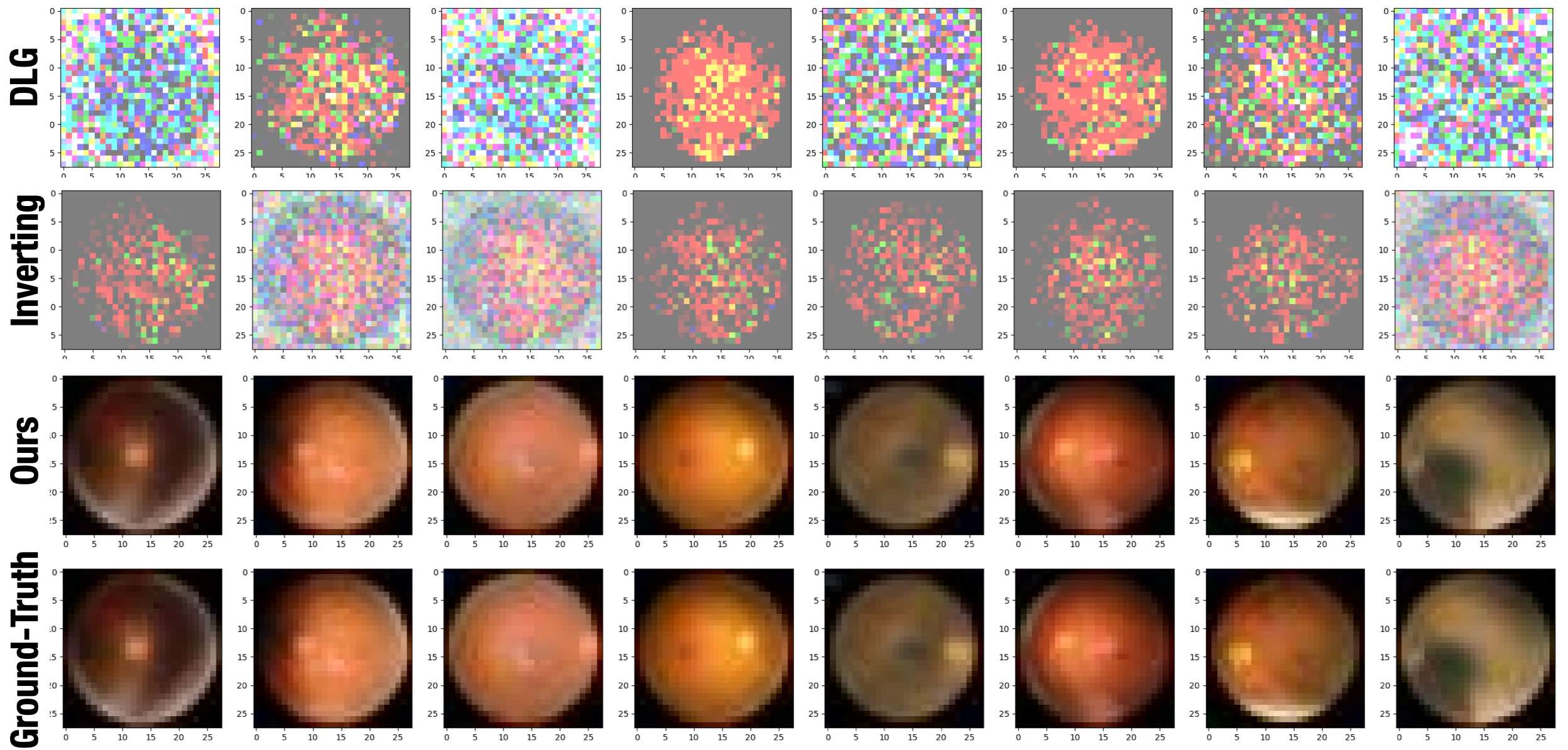}
\caption{Sampled reconstruction results on RetinaMNIST.}
\label{fig:demo_body}
\end{center}
\end{figure}

As the \textbf{LAcc} columns of Table \ref{tab:exp_full_table} show, our attack algorithm reaches $100\%$ accuracy when reconstructing the labels of each single sample in the batch, which conforms to the theoretical guarantee in Theorem \ref{col:optim_bound_mult}. In terms of the MSE and PSNR metrics, our attack algorithm substantially outperforms all the baselines in most test cases. For example, the average PSNRs of our reconstruction results are observed to be larger than $35$ in most cases, which corresponds to highly recognizable reconstruction results for human observers (Fig. \ref{fig:demo_body}). As a comparison, previous attacks tend to produce less recognizable reconstruction results. In the following, we provide more ablation studies to validate the robustness of our proposed attack once the batch has sufficient exclusivity. Due to the space limit on the main text, we omit the full results on all the datasets only if they do not violate the observations we make. The omitted results are all presented in Appendix \ref{sec:app:exp}.

\input{tex/tables/fcn_results.tex}

\begin{figure}[ht]
\begin{center}
\includegraphics[width=0.4\textwidth]{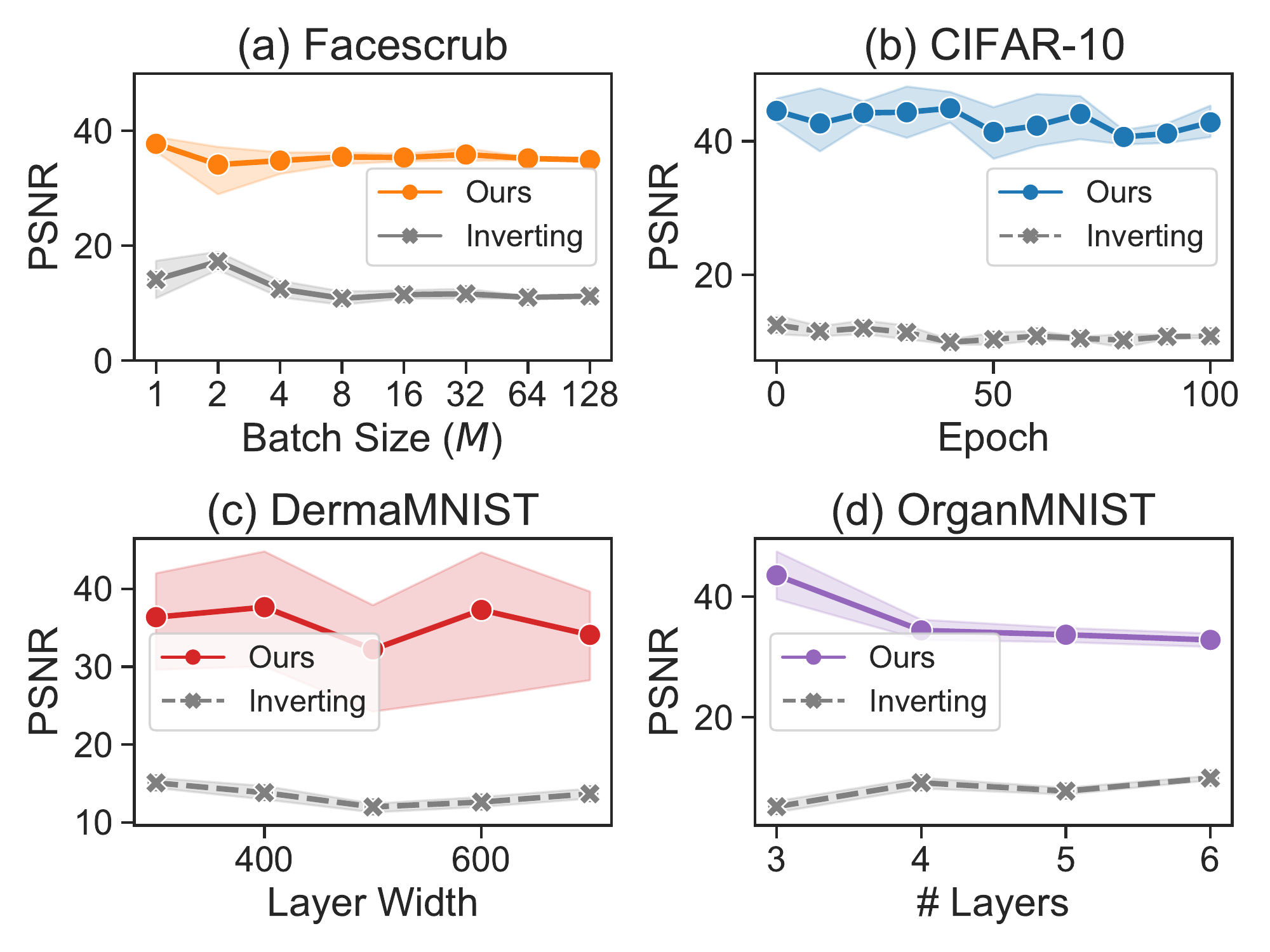}
\caption{The PSNR curve of reconstruction attacks when \textbf{(a)} the batch size, \textbf{(b)} the training epoch, \textbf{(c)} the layer width and \textbf{(d)} the number of layers vary.}
\label{fig:abl_suite}
\end{center}
\end{figure}

\noindent\textbf{Attacks on Partially/Fully Trained Models.} We provide experiments to show the effectiveness of our attack algorithm is not limited to attacking a randomly initialized neural network, but it can also successfully attack partially/fully trained neural networks. Specifically, we train a three-layer fully connected neural network ($d_0$-$512$-$K$) for $100$ epochs, during which the model checkpoints are stored for every $10$ epochs. We conduct our attack and the best baseline \textit{Inverting} on $10$ randomly sampled insecure batches. Fig. \ref{fig:abl_suite}(a) reports the PSNR metrics on CIFAR-10 when the training epoch proceeds from $0$ (i.e., \textit{initial stage}) to $100$ (i.e., \textit{convergence}) with a stride of $10$, where the shaded region reports the $95\%$ confidence interval. As Fig. \ref{fig:abl_suite}(a) shows, the performance of our attack remains stable throughout the whole training process. On CIFAR-10, the MSE of the reconstruction results remain at the $10^{-4}$ error level and the PSNR remains over $40$. Conforming to Theorem \ref{col:optim_bound_mult}, these phenomenons further validate that our attack algorithm works independent from the attack epoch.


\noindent\textbf{Scalability for Realistic Batch Sizes.} To validate the scalability of our proposed attack, we alternatively leverage an auxiliary algorithm in \cite[Section 3.3]{pan21tafa} (referred to as \textit{SOW}) to arbitrarily manipulate the activation pattern of a given input by adding a slight perturbation to the input. We specify the expected activation pattern of each sample in a randomly sampled batch of realistic batch sizes to satisfy the sufficient exclusivity condition. Then, we invoke SOW to generate the perturbations, and conduct our proposed attack on the average gradient of the perturbed batches of size varying from $1$ to $128$ by a multiplier of $2$. Fig. \ref{fig:abl_suite}(b) reports the PSNR metrics of our proposed attack and \textit{Inverting} on Facescrub. We repeat the experiments on $10$ randomly sampled batches, where the shaded part reports the $95\%$ confidence interval of the results. As Fig. \ref{fig:abl_suite}(b) shows, the performance of our attack remains strong when the batch size increases from $1$ to realistic batch sizes like $64$ and $128$ (Fig. \ref{fig:ac1_size_demo_facescrub}). For example, the average PSNR of our attack is $37.8$ and $35.1$ when the batch size is $1$ and $128$ respectively on Facescrub, while the PSNR of Inverting is only $14.1$ and $11.2$. Besides, according to the MSE and PSNR curves, the performance of our proposed attack is almost not correlated with the size of the batch to reconstruct only if the batch stays within the insecure boundary.

\noindent\textbf{Attacks on Different Architectures.} To test our attack on different FCN architectures, we vary the width $d_1$ of the ReLU layer of a 3-layer FCN ($d_0$-$d_1$-$K$) from $300$ to $700$ with a stride of $100$. The corresponding PSNR for $M = 8$ on DermaMNIST is plotted in Fig. \ref{fig:abl_suite}(c). Fixing the layer width as $512$, we also increase the depth of the target FCN ($d$-$512$-$K$) by inserting additional ReLU layers of the same width incrementally to obtain FCNs of $3$-$6$ layers. we report the corresponding PSNR curves on OrganMNIST in Fig. \ref{fig:abl_suite}(d). 

From Fig. \ref{fig:abl_suite}(c)-(d), we observe when the layer width and the number of layers increase, the performance of the learning-based reconstruction attack does not show a clear upward trend, mainly because the gradient-descent-based optimizer is likely to get stuck at a local optimum \cite{Dauphin2014IdentifyingAA} when the learning process converges, which is however distant from the ground-truth results. Consequently, the corresponding PSNR metrics only loosely reflect the intrinsic relation between the model size and the attack effectiveness. Meanwhile, as the PSNR of our attack remains over $20$ in most cases, the improvement of attack performance is also not clear. Nevertheless, a deeper, wider FCN architecture does facilitate data reconstruction attacks according to our analysis: On the one hand, it increases the possibility of a batch to be insecure (Section \ref{sec:impact_factor}). On the other hand, it provides the adversary more scalar equations to determine the data input, which, according to Theorem \ref{col:optim_bound_mult}, lowers down the upper bound on the reconstruction error (Appendix \ref{sec:app:alg}). To alleviate the threats of data reconstruction, one may consider reduce the size of the neural networks especially when the utility requirement is already met.

\begin{figure}[ht]
\begin{center}
\includegraphics[width=0.45\textwidth]{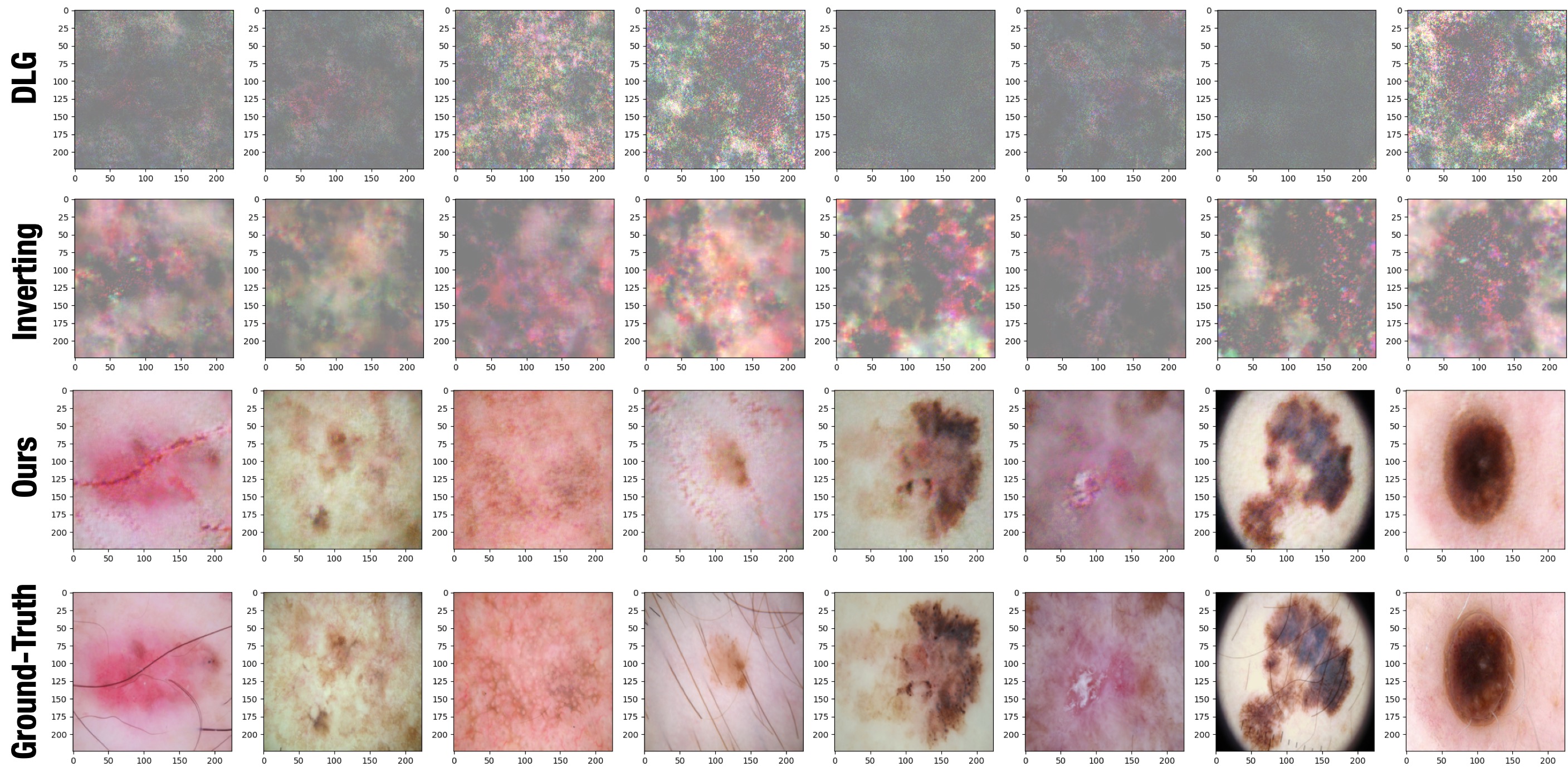}
\caption{Sampled results on the ISIC skin cancer dataset, reconstructed from the average gradient of VGG-13.}
\label{fig:ac2_vgg_demo}
\end{center}
\end{figure}

\noindent\textbf{Hybrid Attacks on CNN-based Classification Models.}
we conduct our hybrid attack on a classical shallow CNN model, i.e., LeNet-5 \cite{Krizhevsky2009LearningML}, and two state-of-the-art deep CNN models, i.e., AlexNet \cite{Krizhevsky2012ImageNetCW} and VGG-13 \cite{Simonyan2015VeryDC}, with three real-world datasets, namely, ImageNet \cite{Russakovsky2015ImageNetLS}, ISIC skin cancer dataset \cite{Gutman2018SkinLA} and Facescrub \cite{Ng2014ADA} (upsampled to $224\times{224}$). The corresponding rows of Table \ref{tab:exp_full_table} report the quantitative performance of our proposed attack and the baseline methods when the batch size is $8$. For better intuition, we also visualize the reconstructed results for VGG-13 on batches from ISIC skin cancer dataset in Fig. \ref{fig:ac2_vgg_demo}. For the omitted visualization on other datasets, please refer to Fig. \ref{fig:app:dcnn_results} in Appendix \ref{sec:app:exp}. 

As we can see from Table \ref{tab:exp_full_table}, our newly proposed hybrid attack on CNN-based classification models outperforms previous attacks, namely, DLG and Inverting, by a non-trivial margin. For example, when reconstructing a batch from ISIC and Facescrub, our proposed attack achieves a PSNR over $20.0$ consistently on all the three representative CNN architectures (with the highest PSNR very close to $30.0$), which conforms to highly recognizable reconstruction results in Fig. \ref{fig:ac2_vgg_demo}. Moreover, by leveraging our proposed attack algorithm on the FCN classifier, we reach $100\%$ accuracy in inferring the labels of each sample in the target batch.

\noindent\textbf{Human Evaluation.} Finally, we measure the reconstruction quality from the perspective of human perception. Specifically, we collect one group of reconstruction results of DLG, Inverting and our attack on the same batch in $8$ test cases when the batch size is $8$. Then we prepare a survey composed of $24$ questions, each of which shows $4$ images ($3$ reconstruction results for the same ground-truth image and the corresponding ground-truth image in a random order) and asks the participant to rank the $4$ images in a decreasing order of recognizability. The study is conducted with $71$ volunteer graduate students. \textbf{This whole study has been approved by our institution's IRB. The approval process is similar to the exempt review in the US, as this study is considered as ``minimal risk'' by IRB staffs.} After collecting the completed surveys, we evaluate the performance of our attack and the baselines in terms of the average discounted cumulative gain (DCG) of the corresponding reconstruction results in each ranking results. Table \ref{tab:human_eval} reports the DCG score of our attack and the baselines on different models averaged over all the participants and the datasets, alongwith the $95\%$ confidence interval. 
Appendix \ref{sec:app:exp} presents more details, with a sample question in Fig. \ref{fig:ac8_sample_question}.    

\input{tex/tables/human_eval.tex}

As Table \ref{tab:human_eval} shows, the human evaluation results are strongly consistent with the performance evaluated with the automatic metrics. For example, on CNNs, our attack always has the second largest DCG score, which is only lower than the ground-truth, for all the target architectures, which conforms to the reported performance in Table \ref{tab:exp_full_table} and indicates the effectiveness of our proposed hybrid extension. More strikingly, on FCNs, our attack even has a higher DCG score under human evaluation compared with the ground-truth, indicating that the reconstructed results from our algorithm are more frequently ranked as the most recognizable than the ground-truth, and conforms to the over $30$ PSNR of our attack on FCNs.

\subsection{Protection Effect inside Secure Boundary} 
\begin{figure}[ht]
\begin{center}

\includegraphics[width=0.4\textwidth]{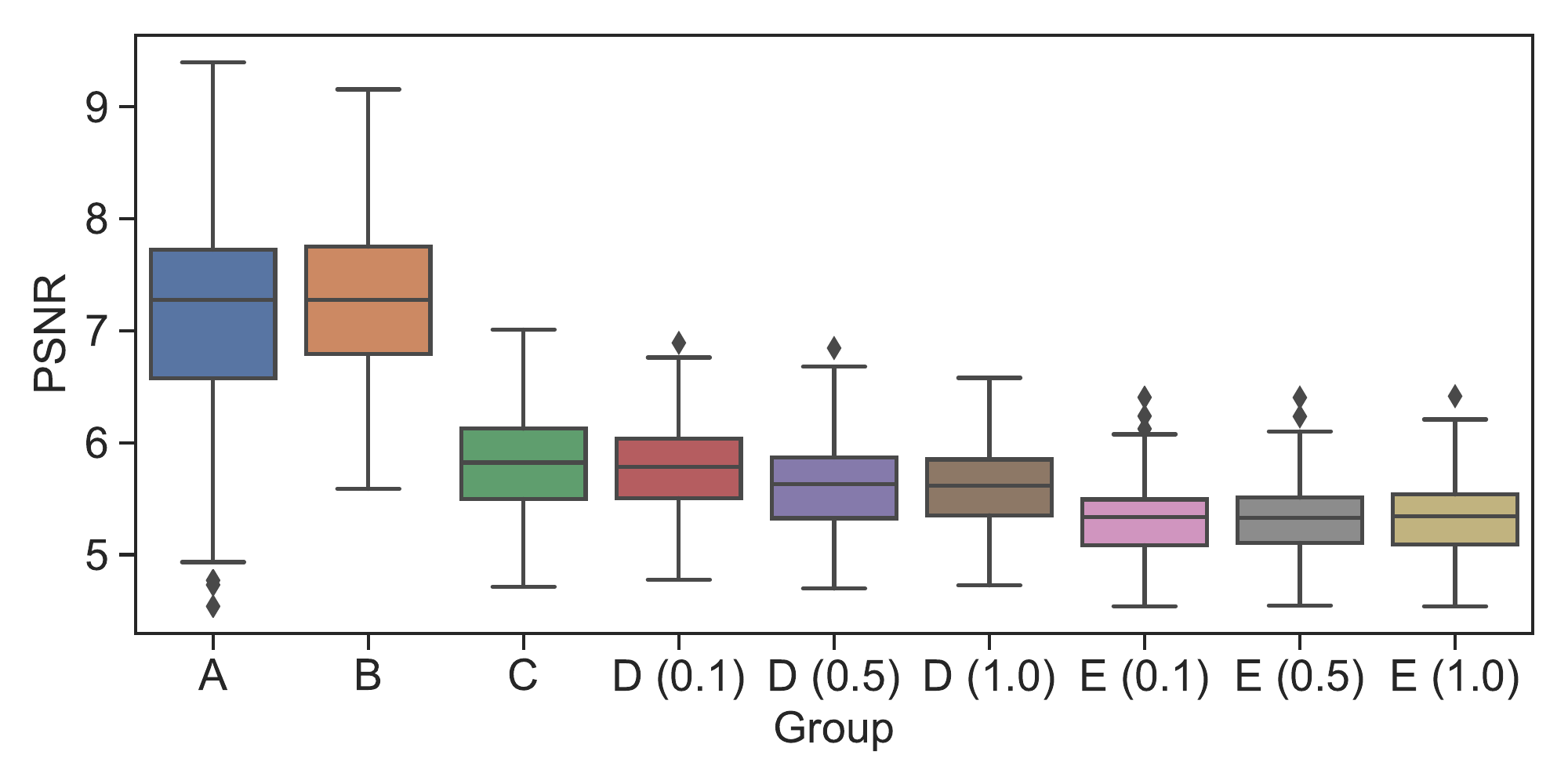}
\caption{The reconstruction quality of \textit{Inverting} on RetinaMNIST when applied on $5$ comparison groups with different architectures or implemented with different defense strategies.}
\label{fig:ac5_defense}
\end{center}
\vspace{-0.25in}
\end{figure}

In this part, we provide preliminary experimental results on how our proposed exclusivity reduction strategy weakens the privacy leakage from gradients. We include differentially-private SGD (DPSGD) \cite{Abadi2016DeepLW} as a potential defense based on gradient obfuscation, orthogonal to our exclusivity reduction strategy which is based on architecture modification. Besides, we further consider a hybrid defense which combines exclusivity reduction with DPSGD. Specifically, we implement the gradient obfuscation procedure of DPSGD as in \cite{Abadi2016DeepLW}, where the gradient clipping constant is set as $1.0$ and the standard deviation $\sigma$ of the Gaussian noise as $0.1$, $0.5$ and $1.0$. In the experiments, we simulate an attacker who leverages the best baseline data reconstruction attack \textit{Inverting} on the average gradient (w/ or w/o obfuscation) of the same batch calculated on the following comparison groups.
\begin{itemize}[leftmargin=*]
    \item \textbf{Group A.} The base FCN ($d_0$-$512$-$K$), i.e., \textit{Base};
    \item \textbf{Group B.} An FCN of the same architecture as in Group A except that a ReLU layer of width $7$ is inserted at the first layer ($d_0$-$7$-$512$-$K$), i.e., \textit{Compression};
    \item \textbf{Group C.} An FCN which shares the same parameters with the model in Group B but has the ReLUs in the first layer removed, i.e., \textit{Compression+w/o ExAN};
    \item \textbf{Group D.} An FCN of the same architecture as in Group B and the gradient is obfuscated with DPSGD, i.e., \textit{Compression+DPSGD} ($\sigma = 0.1, 0.5, 1.0$);
    \item \textbf{Group E.} An FCN of the same architecture as in Group C and the gradient is obfuscated with DPSGD, i.e., \textit{Compression+DPSGD+w/o ExAN} ($\sigma = 0.1, 0.5, 1.0$),
\end{itemize}
where we choose the width of the non-ReLU layer as $7$ because this setting is expected to enhance the privacy of a batch with its size $M \ge 8$ ($=7+1$) according to Corollary 1, which is also a common setting on the maximal size of a batch under attack in previous attacks. For all the five comparison groups, we repetitively conduct the attack on $100$ randomly sampled batches, and collect the average MSE and PSNR as indicators of the reconstruction quality. Fig. \ref{fig:ac5_defense} presents the box-plots of the performance metrics on RetinaMNIST. The omitted results on other datasets are in Appendix \ref{sec:app:alg}.

First, comparing the PSNR on Group A \& B in Fig. \ref{fig:ac5_defense}, we observe the Inverting attack on RetinaMNIST has almost the same performance whether a $7$-unit ReLU layer is inserted into the original model, which indicates the model compression only has a very slight effect in weakening the reconstruction quality. As a comparison, our proposed exclusivity reduction strategy substantially decreases the PSNR of the reconstruction: The PSNR for Group C is $20\%$ lower than the PSNR of Group A \& B. The results imply that exclusivity reduction does play a non-trivial role in weakening the effectiveness of data reconstruction when the compression effect of the shallow layer of width $7$ is left out.  
  
Next, comparing the attack performance on Group C and D, we observe that the attack effectiveness of Inverting is weakened on both groups, which supports that the mitigation strategies via architecture modification or via gradient obfuscation can both alleviate the information leakage from the gradient. Meanwhile, by comparing the decrease in PSNR, we observe that the DPSGD provides as a slightly more effective defense than exclusivity reduction, for which we infer the reason is DPSGD works by directly obfuscating the gradient, the immediate information source exploited by data reconstruction, while our strategy works by reducing the neuron exclusivity, a more in-depth factor which guarantees the non-uniqueness of reconstruction. The orthogonality of these two approaches further inspires us to evaluate a more effective defense which combines our strategy for eliminating the insecure exclusivity state and DPSGD for gradient obfuscation. As the reported performance on Group E shows, this new combination exhibits a larger decrease on the reconstruction quality, while, with regression tests, we observe almost no further trade-off on the normal utility. 

\subsection{Impact Factors on Exclusivity States}
\label{sec:impact_factor}
\begin{figure}[ht]
\begin{center}
\includegraphics[width=0.45\textwidth]{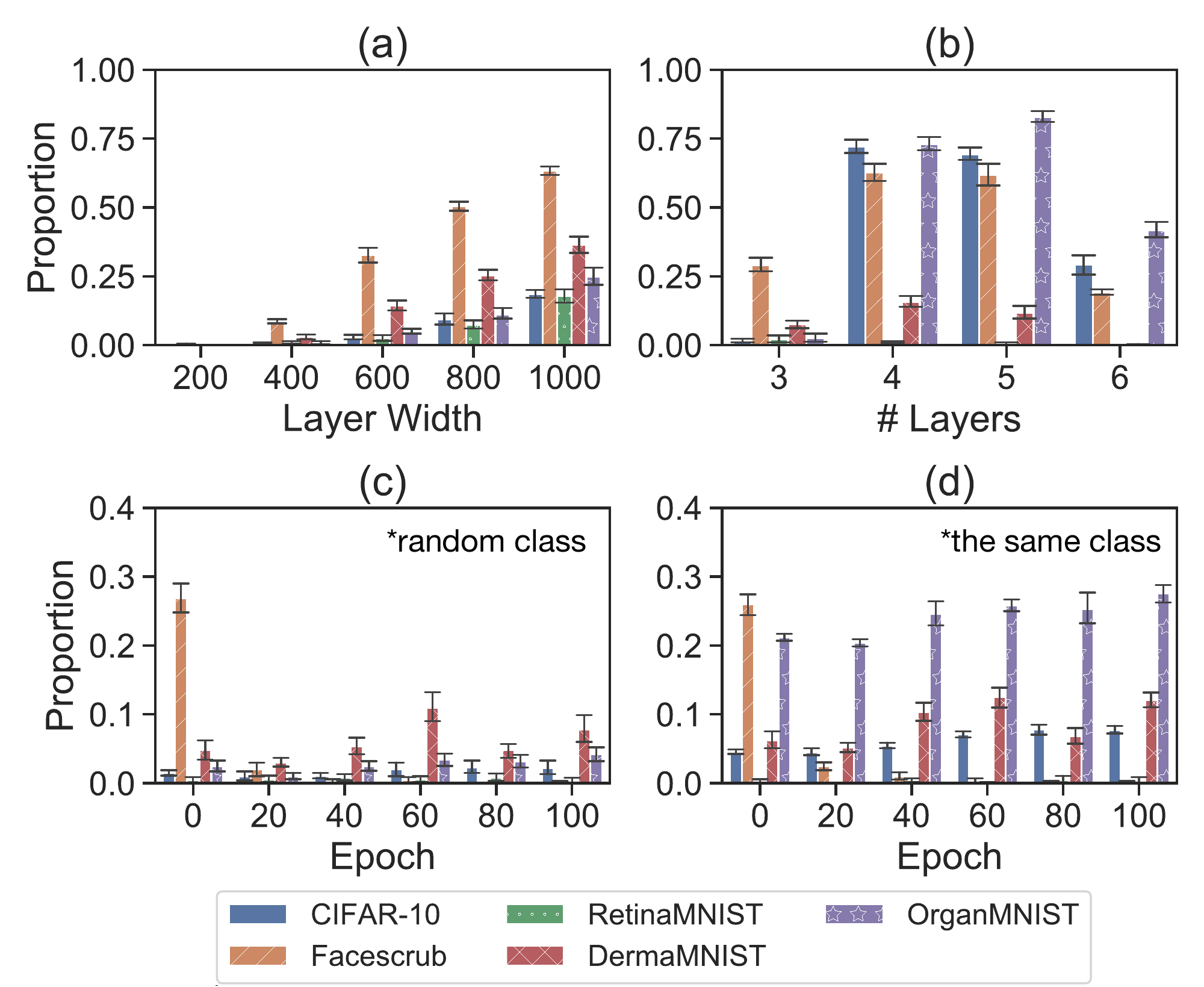}
\caption{The proportion of insecure batches in $1000$ randomly sampled batches under different configurations.}
\label{fig:b4_impact_factor}
\end{center}
\vspace{-0.25in}
\end{figure}

Finally, we empirically study how the layer width, the network depth, the training epoch and the label composition in a batch would influence the statistics of batches which satisfy the sufficient exclusivity condition (i.e., \textit{insecure batch}). Generally, we set the base FCN architecture as a three-layer FCN ($d_0$-$512$-$K$), vary the architecture as specified by the experimental purpose, test the validity of the sufficient exclusivity condition for $1000$ randomly sampled batches of size $8$, and report the proportion of the insecure batches in Fig. \ref{fig:b4_impact_factor}. Specifically, the model configurations in Fig. \ref{fig:b4_impact_factor}(a)-(c) are the same as the ones in Fig. \ref{fig:abl_suite}, while, in Fig. \ref{fig:b4_impact_factor}(d), we report the proportion of valid batches which consist of $8$ samples from the same class, which are averaged over all the classes during the training process, to measure the impact of label composition in a training batch on its exclusivity state.  

As Fig. \ref{fig:b4_impact_factor}(a) shows, on all the five datasets, the width plays a strong impact factor on the proportion of insecure batches. For example, when the layer width is $1000$, the proportion of insecure batches is over $60\%$ on RetinaMNIST, which, in other words, indicates that over $60\%$ batches of size $8$ can be reconstructed with high recognizability only if the average gradient is leaked in this case. From Fig. \ref{fig:b4_impact_factor}(b), we observe that the influence of the network depth on the neuron exclusivity state is complicated. In most cases, the proportion of the insecure batches reaches the maximal when the network depth is $4$ and $5$ but radically decreases when the depth is further enlarged. This may serve as an explanation on our previously reported results in Fig. \ref{fig:abl_suite}(d), where the baseline attacks do not show a clear upward trend when the depth increases. In Fig. \ref{fig:b4_impact_factor}(c), we do not observe a common principle which characterizes the influence of the training epoch on the neuron exclusivity. For example, on Facescrub, the proportion of insecure batches decreases when the training epoch accumulates, while, on DermaMNIST, the proportion first increases and then remains stable. From Fig. \ref{fig:b4_impact_factor}(d), we observe that, on some datasets, the proportion of insecure batches is even higher compared to the case when the batches contain randomly sampled inputs, which conforms to the observed complexity of activation patterns even for samples from the same class \cite{Hanin2019ComplexityOL}, considering the exponentially many possibilities (e.g., $2^{512}$).   

%% file: tex/exp_setting.tex
\subsection{Overview of Evaluation}\label{sec:evaluation_setting}


\noindent\textbf{Datasets.} We provide an overview on the 
$5$ real-world datasets and the corresponding learning tasks in Table \ref{tab:scenarios}. Based on considerations of research ethics, we choose public datasets to construct the data-sensitive scenarios for evaluations. As our attack requires almost no prior knowledge about the datasets, we do think the reported results would faithfully reflect the potential threats to the confidentiality of private training data in the real world. For more details on each scenario, please refer to Appendix \ref{sec:app:datasets}.



\noindent\textbf{Evaluation Protocols.} Following \cite{Zhu2019DeepLF,Geiping2020InvertingG}, we first leverage the Hungarian algorithm \cite{Kuhn1955TheHM} to find the best-matching pairs of reconstructed and ground-truth data inputs according to the pairwise mean square error (MSE). Then we compute the average of the following set of performance metrics over the best-matching pairs. We denote each reconstructed (ground-truth) data input as $\hat{X}_m$ ($X_m$).  

\noindent$\bullet$ \textbf{Mean Square Error (MSE)} measures the L2 difference between the reconstructed input and the ground-truth input, averaged over coordinates. Formally, the MSE metric writes $ \text{MSE}(\hat{X}_m, X_m) = \|\hat{X}_m - X_m\|_2/\dim{\mathcal{X}} $, where $\dim{\mathcal{X}}$ is the dimension of the input space. The MSE is the lower the better.

\noindent$\bullet$ \textbf{Peak Signal-to-Noise Ratio (PSNR)} measures the ratio of the effective information and noises in the reconstructed images, which is also used in \cite{Geiping2020InvertingG}. It formally computes as $
\text{PSNR}(\hat{X}_m, X_m) = -10\times{\log_{10}(\text{MSE}(\hat{X}_m, X_m))}$. It is worth to notice, although PSNR is a derived metric from MSE, it behaves slightly different when being averaged and provides a better perspective on comparing the recognizability of the reconstructed input, especially for the visual scenarios.

Besides, we report the label recovery accuracy, i.e., \textit{LAcc}, which computes the ratio between the number of the labels present in both the ground-truth and the reconstructed label sets with the ground-truth batch size. Moreover, we also visualize the reconstructed results and incorporate human evaluation to better reflect the perceptual reconstruction quality. For more details on other common settings, please refer to Appendix \ref{sec:app:config}.



%% file: tex/tables/fcn_results.tex
\begin{table}[t]
\caption{Comparisons of reconstruction attacks on different scenarios. All statistics are averaged on $10$ controlled repetitive tests, with the best in \textbf{{bold}}.}
\begin{center}

\scalebox{0.55}{
    \begin{tabular}{clccccccccc}
    \toprule
          &       & \multicolumn{3}{c}{\textbf{DLG}} & \multicolumn{3}{c}{\textbf{Inverting}} & \multicolumn{3}{c}{\textbf{Ours}} \\
\cmidrule{3-11}          &       & \textbf{MSE} & \textbf{PSNR} & \textbf{LAcc} & \textbf{MSE} & \textbf{PSNR} & \textbf{LAcc} & \textbf{MSE} & \textbf{PSNR} & \textbf{LAcc} \\
    \midrule
    \multirow{5}[2]{*}{\textbf{FCN}} & \textbf{CIFAR-10} & 0.503 & 8.75  & 0.475  & 0.296  & 12.50  & 0.775  & \textbf{0.001 } & \textbf{48.12 } & \textbf{1.000 } \\
          & \textbf{RetinaMNIST}  & 1.102 & 4.48  & 0.500  & 0.993  & 4.97  & 0.513  & \textbf{0.030 } & \textbf{19.88 } & \textbf{1.000 } \\
          & \textbf{DermaMNIST}  & 0.15  & 10.87  & 0.450  & 0.095  & 17.11  & 0.775  & \textbf{0.005 } & \textbf{41.42 } & \textbf{1.000 } \\
          & \textbf{OrganMNIST}  & 0.565 & 7.77  & 0.375  & 0.263  & 12.95  & 0.775  & \textbf{0.012 } & \textbf{43.56 } & \textbf{1.000 } \\
          & \textbf{Facescrub} & 0.604 & 6.94  & 0.475  & 0.360  & 11.59  & 0.588  & \textbf{0.002 } & \textbf{35.48 } & \textbf{1.000 } \\
    \midrule
    \multicolumn{1}{c}{\multirow{3}[2]{*}{\textbf{LeNet-5\newline{}}}} & \textbf{ImageNet} & 0.496  & 13.46  & 0.375  & 0.213  & 13.26  & 1.000  & \textbf{0.046 } & \textbf{19.52 } & \textbf{1.000 } \\
          & \textbf{ISIC} & 0.438  & 9.68  & 0.375  & 0.086  & 17.31  & 1.000  & \textbf{0.071 } & \textbf{24.93 } & \textbf{1.000 } \\
          & \textbf{Facescrub} & 0.699  & 7.77  & 0.500  & 0.245  & 12.73  & 0.625  & \textbf{0.007 } & \textbf{28.88 } & \textbf{1.000 } \\
    \midrule
    \multicolumn{1}{c}{\multirow{3}[2]{*}{\textbf{AlexNet \newline{}}}} & \textbf{ImageNet} & 0.513  & 9.06  & 0.375  & 0.370  & 10.57  & 0.875  & \textbf{0.229 } & \textbf{12.79 } & \textbf{1.000 } \\
          & \textbf{ISIC} & 0.247  & 12.51  & 0.500  & 0.093  & 17.32  & 0.875  & \textbf{0.018 } & \textbf{24.90 } & \textbf{1.000 } \\
          & \textbf{Facescrub} & 0.677  & 7.86  & 0.625  & 0.298  & 11.59  & 0.875  & \textbf{0.037 } & \textbf{20.48 } & \textbf{1.000 } \\
    \midrule
    \multicolumn{1}{c}{\multirow{3}[2]{*}{\textbf{VGG-13 \newline{}}}} & \textbf{ImageNet} & 0.404  & 10.11  & 0.375  & 0.292  & 11.89  & 1.000  & \textbf{0.087 } & \textbf{17.55 } & \textbf{1.000 } \\
          & \textbf{ISIC} & 0.173  & 14.20  & 0.625  & 0.114  & 16.17  & 1.000  & \textbf{0.006 } & \textbf{28.14 } & \textbf{1.000 } \\
          & \textbf{Facescrub} & 0.255  & 12.01  & 0.125  & 0.212  & 13.25  & 0.875  & \textbf{0.007 } & \textbf{29.53 } & \textbf{1.000 } \\
    \bottomrule
    \end{tabular}%
}
\end{center}
\label{tab:exp_full_table}
\end{table}

%% file: tex/tables/human_eval.tex
\begin{table}[ht]
  \centering
  \caption{Comparison of different attacks in terms of perceptual reconstruction quality in terms of discounted cumulative gain (DCG).}
  \scalebox{0.75}{
    \begin{tabular}{lcccc}
    \toprule
          & \textbf{DLG} & \textbf{Inverting} & \textbf{Ours} & \textbf{Ground-Truth} \\
    \midrule
    \textbf{FCN} & 0.432 $\pm$ 0.002 & 0.503 $\pm$ 0.003 & \textbf{0.83 $\pm$ 0.01} & 0.80 $\pm$ 0.01 \\
    \textbf{LeNet-5} & 0.448 $\pm$ 0.004 & 0.487 $\pm$ 0.003 & 0.72 $\pm$ 0.02 & \textbf{0.91 $\pm$ 0.01} \\
    \textbf{AlexNet} & 0.436 $\pm$ 0.004 & 0.502 $\pm$ 0.005 & 0.630 $\pm$ 0.006 & \textbf{0.993 $\pm$ 0.005 } \\
    \textbf{VGG-13} & 0.442 $\pm$ 0.006 & 0.496 $\pm$ 0.003 & 0.70 $\pm$ 0.01 & \textbf{0.93 $\pm$ 0.01} \\
    \bottomrule
    \end{tabular}}%
  \label{tab:human_eval}%
\end{table}%

%% file: tex/discussion.tex
\section{Discussions}\label{sec:discussion}


\begin{figure}
\begin{center}
\includegraphics[width=0.45\textwidth]{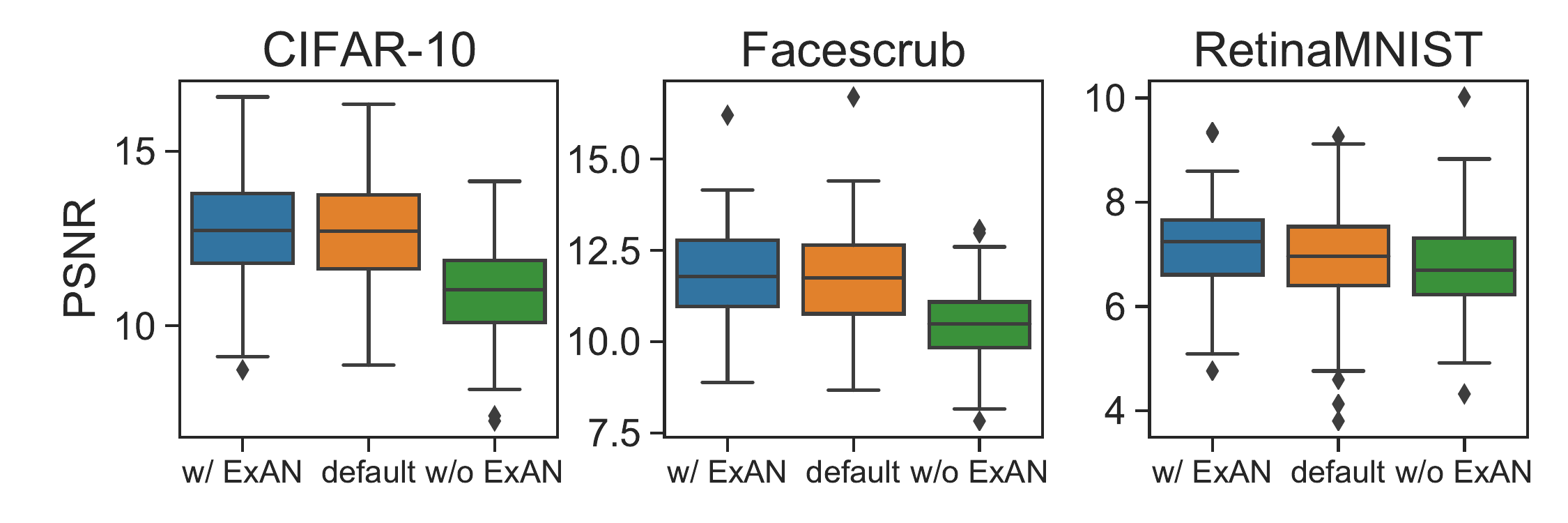}
\caption{Impact of \textit{sufficient exclusivity} (i.e., \textit{w/ ExAN}) and \textit{lack of exclusivity} (i.e., \textit{w/o ExAN}) conditions on the quality of data reconstruction, where the \textit{default} columns collect the results on a randomly sampled batches with no control on its exclusivity state.}
\label{fig:b4_exan_asr}
\end{center}
\vspace{-0.3in}
\end{figure}

\noindent{\textbf{On the Remaining Exclusivity States.}} We further explore how the neuron exclusivity state of a batch would influence the attack effectiveness in general. Specifically, with the aid of the SOW algorithm \cite{pan21tafa}, we prepare the following three comparison groups of batches to attack: \textbf{A.} $100$ randomly sampled batch of size $8$ from the original dataset (i.e., \textit{default}); \textbf{B.} slightly perturbed versions of batches in Group A such that each batch satisfies the condition of sufficient exclusivity (i.e., \textit{w/ ExAN}); \textbf{C.} slightly perturbed versions of batches in Group A such that all the samples in the same batch have no ExAN with one another. Then we conduct the \textit{Inverting} attack on all the $100$ batches from the three comparison groups respectively. Fig. \ref{fig:b4_exan_asr} presents the box plots of the PSNR of \textit{Inverting} on three datasets, with the omitted results in Appendix \ref{sec:app:alg}. As Fig. \ref{fig:b4_exan_asr} shows, in most cases, we observe that the attack performance decreases in the following order of the comparison groups: \textit{w/ ExAN} $>$ \textit{default} $>$ \textit{w/o ExAN}. For example, on Facescrub, the average PSNR is respectively $11.83$, $11.55$, and $10.47$ for the \textit{w/ ExAN}, the \textit{default} and the \textit{w/o ExAN} comparison group. Although the performance margin of \textit{Inverting} between the default group and the \textit{w/ ExAN} group is not as substantial as that between the default group and the w/o ExAN group, we should notice \textit{Inverting} does not exhibit the optimal attack effectiveness on insecure batches. In fact, our constructed attack algorithm indicates, the attacker can achieve a much higher reconstruction quality (e.g., with PSNR over $30$) on the insecure batches. Combining these results, we expect ExAN as a promising metric for understanding and measuring the data leakage from the gradient. Yet, there lacks rigorous statements whether an effective data reconstruction attack can be constructed for the remaining cases as in Section \ref{sec:atk_case_a}, or whether the impossibility of unique reconstruction can be proved as in Section \ref{sec:def_case_b}, which is left as an open question for future research.

\noindent{\textbf{Data Reconstruction vs. Model  Extraction.}} As pointed out in \cite{Jagielski2020HighAA}, model extraction  becomes less feasible when the model is expansive (i.e., the model contains a layer with a higher output dimension than the input dimension), while, under the same condition, data reconstruction attack in turn becomes stronger, according to our analysis. It is mainly because, the information exposed to model extraction attacks is a number of data inputs (i.e., queries) and their predictions, from which he/she wants to recover the model parameters. Therefore, model extraction on an expansive model has to recover more unknown variables than either the input dimension or the prediction dimension, which becomes an issue. In contrast, the information exposed to data reconstruction attacks is the gradient, the information of which grows when the model becomes larger. When the model is expansive, the gradient information accessible to the adversary is sufficiently more than the dimension of the unknown inputs the adversary wants to solve, which further facilitates data reconstruction.

\noindent\textbf{Limitation \& Future Directions.} To further improve the reconstruction accuracy of our analytic attack, future works may consider design a solution refinement procedure based on non-convex optimization techniques \cite{Jagielski2020HighAA}, e.g., by using the solved solution from the linear equation solver as the initial guess and then refining the solution iteratively by gradient descents on the gradient matching objective. Besides, our current work mainly characterizes the defensive effectiveness of exclusivity reduction with the theory of linear equation systems. As a promising future work, one may consider extend the analytic results to the language of differential privacy. Finally, future works may also study the role of neuron exclusivity states in other gradient-based privacy attack classes. For example, our proposed exclusivity reduction may also weaken the effectiveness of gradient-based property inference attacks \cite{Melis2019ExploitingUF}, because the gradient information after exclusivity reduction can also correspond to many other mini-batches which do not share the same global property with the original mini-batch, which can therefore obfuscate the attacker's inference.  




%% file: tex/cls.tex
\section{Conclusion}
In this paper, we provide the first analytic study which explores the security boundary of data reconstruction from gradient via the lens of neuron exclusivity states. Specifically, we determine and prove the boundary condition of insecure exclusivity states by constructing an attack algorithm with guaranteed accuracy. Moreover, we prove the impossibility of unique reconstruction for the exclusivity states satisfying the lack of exclusivity condition. With our proposed simple yet effective exclusivity reduction strategy as a preliminary step, we hope our study would arouse more research interests and efforts in investigating and strengthening the privacy properties of model gradient via its intrinsic interaction with the underlying mechanism of deep learning.    

\input{tex/ack}

%% file: tex/ack.tex
\section*{Acknowledgments}
We would like to thank the anonymous reviewers for their constructive comments and input to improve our paper. This work was supported in part by National Natural Science Foundation of China (61972099, U1836213,U1836210, U1736208), and Natural Science Foundation of Shanghai (19ZR1404800). Min Yang is a faculty of Shanghai Institute of Intelligent Electronics \& Systems, Shanghai Institute for Advanced Communication and Data Science, and Engineering Research Center of CyberSecurity Auditing and Monitoring, Ministry of Education, China. 

%% file: tex/appendix.tex
\appendix 
\renewcommand\thefigure{\thesection.\arabic{figure}}  
\renewcommand\thealgorithm{\thesection.\arabic{algorithm}}
\renewcommand\thetable{\thesection.\arabic{table}}

\setcounter{figure}{0} 
\setcounter{table}{0}
\setcounter{algorithm}{0}

\input{tex/app/proofs.tex}

\input{tex/app/datasets.tex}
\input{tex/app/implementation.tex}

\input{tex/app/eval.tex}

%% file: tex/app/proofs.tex
\section{Omitted Proofs for Analytic Results}\label{sec:app:proof}

\noindent{\textbf{Proof for Proposition \ref{prop:reconstruct_D}.}}
For convenience, we denote the $j$-th element of $D_i^{m}$ as $\alpha_{i,j}^{m}$, i.e., the activation state of the $j$-th neuron at the $i$-th layer when $X_m$ is the input. Formally, the exclusive activation of a neuron is expressed as: $\alpha_{i,j}^{m}$ takes the value $1$ for and only for a certain sample $X_m$. For intuition, readers may refer to Fig. \ref{fig:activ_pattern} as an illustrative example.
\begin{itemize}
\item \textit{Initial Step}: As a by-product of solving $\overline{g}_{c}^{m}$ and the assumed exclusivity, we already recovered at least two exclusive elements in $D_H^{m}$ for each input $X_m$.

\item \textit{Recurrent Step}: Next, we consider the gradient equation w.r.t. $W_{H-1}$.

\begin{equation}
\overline{G}_{H-1} = \frac{1}{M}\sum_{m=1}^{M}\sum_{c=1}^{K}\overline{g}_{c}(D_{H-1}^{m}...W_0X_m)([W_H]_c^{T}D_H^{m})
\end{equation}

Then, we expand it explicitly to individual scalar equations.
\begin{align}
M[\overline{G}_{H-1}]_{ij} = \sum_{m=1}^{M}\sum_{c=1}^{K}\overline{g}_{c}\alpha_{H-1, i}^{m}f_{H-2,i}^{m}[W_{H}]_{jc}\alpha_{H,j}^{m} \nonumber\\ := \sum_{m=1}^{M}C_{ij}^{m}\alpha_{H-1, i}^{m}\alpha_{H,j}^{m}
\end{align}

In the last line, we use the $C_{ij}^{m}$ to replace the multiplier (which is non-zero almost surely in our threat model). The following is the key of the recurrent step. As $\{D_{H}^{m}\}_{m=1}^{M}$ have at least one exclusive nonzero position to each other, the terms in the summation above therefore have at most one non-vanishing term for this ExAN, indexed by e.g., $j$, which can be found based on the knowledge of $\{D_{H}^{m}\}_{m=1}^{M}$. In fact, the $j$-th column of $\overline{G}_{H-1}$, i.e., $[C_{ij}^{m}\alpha_{H-1,i}^{m}]$, immediately gives the diagonal terms of $D_{H-1}^{m}$, if we simply check the non-zero positions of $[\overline{G}_{H-1}]_{:, j}$. Similarly, with the solved $\{D_{H-1}^{m}\}_{m=1}^{M}$, the procedure can be done for the $(H-2)$-th layer, and so on, until the input layer. 
\end{itemize}

\noindent{\textbf{Proof for Theorem \ref{col:optim_bound_mult}}.} This case corresponds to the situation when the gradient equation system is under-determined, i.e., the number of equations is smaller than the number of variables. We denote the total derivative operator $A := (\nabla_{W_0}{\ell}, \hdots, \nabla{W_{H}}{\ell})$, where $\nabla_{W_i}{\ell}(X_1, \hdots, X_M) = \frac{1}{M}\sum_{m=1}^{M}\nabla_{W_i}{\ell^{m}(X_m)}$ (Here, ${\ell^{m}(X_m)}$ is defined similarly to the average loss while the accumulated activation patterns $(D_1, \hdots, D_H)$ are replaced by the $m$-th sample's own activation pattern $(D_1^m, \hdots, D_{H}^{m})$). Therefore, the (under)-determined gradient equation writes $A(X_1, \hdots, X_M) = (G_0, \hdots, G_H) := b$, which has the ground-truth data inputs $X^{*} := (X_1^{*}, \hdots, X_M^{*})$ as the least-square-error (LSE) solution. Then, we need to consider, when the attacker is only provided with an underdetermined equation system, i.e., $(A+\Delta{A})X = b+\Delta{b}$, how the corresponding LSE solution $X := (X_1, \hdots, X_M)$ is perturbed. We introduce the following lemma. 
\begin{lma}[Theorem 5.7.1\cite{Golub1996MatrixC}]
Suppose $\text{rank}(A) = m \ge n$ and that $A\in\mathbb{R}^{m\times{n}}$, $\Delta{A}\in\mathbb{R}^{m\times{n}}$,
$0\neq{b}\in\mathbb{R}^{m}$, and $\Delta{b}\in\mathbb{R}^{m}$ satisfy
$\epsilon = \max{\epsilon_A, \epsilon_b} < \lambda(A)$, where $\epsilon_A = \|\Delta{A}\|_2/\|A\|_2$ and $\epsilon_b = \|\Delta{b}\|_2/\|b\|_2$. If $x$ and $\hat{x}$ are minimum norm solutions that satisfy $Ax = b$ and $(A+\Delta{A})\hat{x} = b + \Delta{b}$, then
\begin{equation}
    \frac{\|\hat{x} - x\|_2}{\|x\|_2} \le \text{cond}(A)(\epsilon_A\min\{2, n-m+1\}+\epsilon_b) + O(\epsilon^2),
\end{equation}
\end{lma}
When the perturbation $\delta:= \|\Delta{A}\|_2/\|A\|_2 < \lambda(A)$ (i.e., the smallest singular value of $A$), we have $\|X-X^{*}\|_2/\|X^*\|_2 < {2\delta}\text{cond}(A)<{2\sum_{i=0}^{H}\delta_i\text{cond}(\nabla_{W_i}{\ell})}$, where $\delta_{i} := \|\Delta{A_i}\|_2/\|\nabla_{W_i}{\ell}\|_2$ and $\Delta{}A_i$ is the perturbation added to the $i$-th layer. First, we consider the perturbation condition to estimate $\delta_{i}$. For the $i$-th layer, the condition requires $\delta_i<\lambda(\nabla_{W_i}{\ell})$. Considering the underdetermined equation system built by the attacker, the perturbation $\Delta{A_i}$ should cancel out the rows of $\nabla_{W_i}{\ell}$ where the gradient is not captured, i.e., the $\|\Delta{A_i}\|_2/\|\nabla_{W_i}{\ell}\|_2 = (1-\beta(\overline{G}_i))$ almost surely (where $\overline{G}_i$ is the gradient at the $i$-th layer captured by the attacker). Next, we apply the following lemma from \cite{AllenZhu2019ACT} to estimate the singular value of $A$, 
\begin{lma}[Theorem 3 \cite{AllenZhu2019ACT}]
For every $i=0, \dots, H$, with probability $\ge 1 - \exp{-\Omega(\sqrt{d_id_{i+1}}/\text{poly}(M, H, \epsilon_i^{'-1}))}$, it satisfies , and every $W_i$ with $\|W_i - W_i^{(0)}\|_2 \le \frac{1}{\text{poly}(M, H, \epsilon_i^{'-1})}$,
\begin{equation}
   \Omega(\frac{\sqrt{d_id_{i+1}}\epsilon_i^{'}}{M\dim{\mathcal{X}}}) \le \|A\|_F^{2} \le O(\frac{\sqrt{d_id_{i+1}}M}{\dim{\mathcal{X}}}).
\end{equation}
\end{lma}
In other words, the smallest and the largest singular values of $A$ are controlled by the two ends of the inequality above. Therefore, the requirement above is reduced to $\delta= (1-\beta(\overline{G}_i)) < \lambda(\nabla_{W_i}{\ell}) = \epsilon^{'}_i{\frac{\sqrt{d_id_{i+1}}}{M\dim{\mathcal{X}}}}$ almost surely.

Using the two estimates in the lemma above, we can further upper bound the conditional number $\text{cond}(\nabla_{W_i}{\ell}) := \Lambda(\nabla_{W_i}{\ell})/\lambda(\nabla_{W_i}{\ell}) < O(\frac{M^2}{\epsilon_i^{'}})$, where $\Lambda(\cdot)$ is the largest singular value. Finally, by inserting the estimations of $\text{cond}(\nabla_{W_i}{\ell})$ and $\delta_i$ into the original bound and replacing $M/\epsilon_i^{'}$ with a new constant $\epsilon_i$, we have $\|X-X^{*}\|_2/\|X^*\|_2 < O({{M}\sum_{i=0}^{H}\epsilon_i(1-\beta(\overline{G}_i))})$, if for all $i\in\{0, \hdots, H\}$, $1-\beta(\overline{G}_i) < \epsilon_i{\frac{\sqrt{d_id_{i+1}}}{M\dim{\mathcal{X}}}}$. Expanding and moving $\|X^{*}\|_2$ to RHS gives the final form in Theorem \ref{col:optim_bound_mult}.

\noindent{\textbf{Proof for Theorem \ref{thm:impossible_reconstr} and Corollary \ref{col:exclusivity_reduction}.}}
We prove the impossibility of unique reconstruction by directly constructing the linear space $\mathcal{Q}$ where every translation $\Delta\in\mathcal{Q}$ satisfies Eq. (\ref{eq:weight_inv}) \& (\ref{eq:bias_inv}). To construct the perturbation $\Delta \in \mathbb{R}^{d_0 \times {M}}$, we only need to consider solve the following equation system.$
\begin{cases}
A\Delta^T = 0 \\ W_0\Delta = 0,
\end{cases}$
where $A = [{\alpha}_1^T, \dots, {\alpha}_M^T]\in\mathbb{R}^{d_1\times{
M}}$ and ${\alpha}_m = \sum_{c=1}^{K}\overline{g}_c^{m}([W_H]_c^{T}D_H^m \hdots W_1D_1^m)$. It is easy to see, for any $\Delta$ satisfying the second equation above, we always have $W_0(X_m + \Delta_m) = W_0X_m$, which guarantees the gradients w.r.t. each $(b_i)_{i=0}^{H}$ and each $(W_i)_{i=1}^{H}$ to be invariant. Meanwhile, to satisfy the first equation guarantees the gradients w.r.t. $W_0$ to be invariant. In the following, we show the solution set of the equation system above itself is a linear space of dimension $M\times(d_0 - d_1)$.

First, we consider the equation $W_0\Delta = 0$. When $d_1 < d_0$, this equation has its solution written as $\Delta = (I - W_0^{\dagger}W_0)Q$, where $W_0^{\dagger}$  is the Moore-Penrose (MP) (pseudo-)inverse and $Q$ is an arbitrary matrix in $\mathbb{R}^{d_0\times{M}}$. Denote the projection operator $P_{0} := I - W_0^{\dagger}W_0$. Inserting the above equation into the first equation $A\Delta^T = 0$, we obtain the following constraint on $\tilde{Q} (:= Q^T)$: $A\tilde{Q}P_0^T = 0$. Next, we utilize the following results from \cite{Magnus1999MatrixDC}.
\begin{lma}[Theorem 2.13\cite{Magnus1999MatrixDC}]
A necessary and sufficient condition for the matrix equation $AXB=C$ to have a solution is that $AA^{\dagger}CB^{\dagger}B = C$, in which case the general solution is $X = A^{\dagger}CB^{\dagger} + Q - A^{\dagger}AQBB^{\dagger}$. 
\end{lma}
In our context, for the equation $A\tilde{Q}P_0^T = 0$, we set $C=0$ in the above lemma, which states the equation always has infinitely many solutions written in $\tilde{Q} = Q - A^{\dagger}AQ(P_0^TP_0^{T\dagger})^T$, where $Q$ is an arbitrary vector in $\mathbb{R}^{d_0\times{m}}$. Thus, we have $\Delta = P_0(Q - A^{\dagger}AQ(P_0^TP_0^{T\dagger}))^T$ for an arbirary $Q\in\mathbb{R}^{d_0\times{m}}$, which, as can be easily checked, forms a linear space $\mathcal{Q}$. Finally, as the projection operator $P_0$ projects the $\mathbb{R}^{m\times{d_0}}$ to a subspace of dimension $m\times{(d_0 - d_1)}$, we have $\dim{\mathcal{Q}} = m\times{(d_0 - d_1)}$.

Next, we show there exists a perturbation subspace $\mathcal{Q}$ such that for any $\Delta\in\mathcal{Q}$, the gradient equation becomes identical for $X$ and $X+\Delta$, which in other words implies the impossibility of unique reconstruction from the gradient equation as the only information source. In this part, we further analyze the property of the perturbation subspace to answer how large such a perturbation can be. As a typical scenario, we estimate the upper bound of $\max \frac{1}{M}\sum_{i=1}^{M}\|\Delta_i\|_2^2$ where $\Delta$ satisfies the above equation system and respects the common box constraint on an image input, i.e., $X + \Delta \in [-1, 1]^{M \times {d_0}}$.

Denote the null space of $W_0$ as $W_0^{\perp} = \text{span}({e}_1, \hdots, {e}_{d_0 - d_1})$, where $({e}_{j})_{j=1}^{d_0-d_1}$ forms the orthogonal basis of $W_0^{\perp}$. Besides, we denote the remaining orthogonal basis as $\{{e}_{d_0-d_1+1}, \dots, {e}_{d_0}\}$. We also denote the basis transformation matrix as $T = [{e}_{1}, \dots, {e}_{d_0}]$. As $\Delta \in W_0^{\perp}$, we represent $\Delta_{i} = \sum_{j=1}^{d_0 - d_1}\delta_{ij}{e}_{j}$. Also with the orthogonal basis of the null space, we reformulate the box constraint $X + \Delta \in [-1, 1]^{M \times {d_0}}$ as an inequality $-\mathbf{1}_{d_0} \preceq X_i + \Delta_{i} \preceq \mathbf{1}_{d_0}$ ($i = 1, \dots, M$). Applying the projection operator $P_0$ related with $W_0^{\perp}$ to both sides of the inequality, we have $-P_0\mathbf{1}_{d_0} \preceq{P_0X_i+\Delta_i} \preceq{P_0\mathbf{1}_{d_0}}$ (note $P_0\Delta_i = \Delta_i$), which gives $-|P_0|\mathbf{1}_{d_0}-P_0X_i \preceq \Delta_i \preceq |P_0|\mathbf{1}_{d_0} - P_0X_i$, where $|\cdot|$ denotes the elementwise absolute on the matrix. Similarly, applying the basis transformation matrix to the inequality, we have ${-|T||P_0|\mathbf{1}_{d_0} - TP_0X_i}\preceq{}T\Delta_i \preceq{|T||P_0|\mathbf{1}_{d_0} - TP_0X_i}$. The inequality is therefore transformed to another set of box constraints $\delta_{ij} \in [-a_{ij}, b_{ij}]$ ($i = 1, \dots, M, j =1, \dots, d_0 - d_1$), where $a_{ij} := [|T||P_0|\mathbf{1}_{d_0}+TP_0X_i]_j$ and $b_{ij} := [|T||P_0|\mathbf{1}_{d_0}-TP_0X_i]_j$.

Then, our problem reduces to estimate the upper bound of $\sum_{i = 1}^{M}(\sum_{j=1}^{d_0 - d_1} \delta_{ij}^{2})^{1/2}$, where $(\delta_{ij})$ satisfy the interval constraints $\delta_{ij} \in [-a_{ij}, b_{ij}]$ and the first matrix equation $A\Delta^{T} = 0$. Inserting the orthogonal basis representation of $\Delta$ into the equation, we have $\sum_{i=1}^{M}\sum_{j=1}^{d_0 - d_1}\delta_{ij}({\alpha}_{i}\otimes{e}_{j}) = 0$, which can be reformulated as the following linear equation w.r.t. $\delta := (\delta_{ij})_{i=1,j=1}^{M, d_0 - d_1}$:
\begin{equation}
      (A\otimes{E})\text{vec}(\delta) = 0
\end{equation}
where $A = [{\alpha}_1^T, \dots, {\alpha}_M^T]\in\mathbb{R}^{d_1\times{M}}$ and $E = [e_1^{T}, \dots, e_{d_0 - d_1}^{T}] \in \mathbb{R}^{d_0 \times{(d_0 - d_1)}}$. As $\text{rank}(A\otimes{E}) = \text{rank}(A)\text{rank}(E) = d_1(d_0 - d_1) < M(d_0 - d_1)$, the linear vector equation above always have infinitely many non-trivial solutions. Denote the projection operator w.r.t. $A\otimes{E}$ as $P_1 = I - (A\otimes{E})^{\dagger}A\otimes{E} = I - (A^{\dagger}A\otimes{E}^{\dagger}E) = I - (A^{\dagger}A\otimes{I_{d_0 - d_1}})$ (as the matrix $E$ formed by the orthogonal basis is of full column rank). With the above definition, the general solution of $A\Delta^{T} = 0$ is written as $P_1{q}$, where ${q}\in\mathbb{R}^{M(d_0 - d_1)}$ satisfies the interval constraints $[-a_{ij}, b_{ij}]$. As the norm of the perturbation $\sum_{i=1}^{M}\sum_{j=1}^{d_0 - d_1}\delta_{ij}^{2}$ is equal to $\|P_1{q}\|_2^{2} = {q}^{T}P_1^TP_1{q}$, a quadratic function with the critical point at ${q} = 0$ with a positive curvature (as $P_1^{T}P_1 \succ 0$), we therefore assert that the maximum norm solution is taken at the boundary points of the interval constraints. Formally, it gives $\max_{{q}}\|P_1{q}\|_2^2 = \|P_1{q}^{*}\|_2^{2}$, where $({q}^{*})_{ij} = \max\{a_{ij}, b_{ij}\} = \max\{([|T||P_0|\mathbf{1}_{d_0}+TP_0X_i]_j, [|T||P_0|\mathbf{1}_{d_0}-TP_0X_i]_j\} \ge |TP_0X_i|_j$. Denote ${\eta}_i = |TP_0X_i|$ and therefore ${q}^{*} = {\eta}_1 \oplus \dots \oplus {\eta}_{M}$. Denote $Y = [{\eta}_1^{T}, \dots, {\eta}_1^{T}] \in \mathbb{R}^{(d_0 - d_1)\times{M}}$. Finally, we have $\|P_1{q}^{*}\|_2^2 = {q}^{*T}P_1^TP_1{q}^{*} = {q}^{*T}P_1{q}^{*} = \|{q}^{*}\|_2^{2} - {q}^{*T}(A^{\dagger}A\otimes{I_{d_0 - d_1}}){q}^{*} = \|{q}^{*}\|_2^{2} - \text{Tr}(A^{\dagger}AY^{T}Y) = \sum_{i=1}^{M}\|{\eta}_i\|_2^2 - \text{Tr}(A^{\dagger}AY^{T}Y)$, where the second equality comes from the fact that the projection operator $P_1$ is symmetric and idempotent. Corollary \ref{col:exclusivity_reduction} is immediate as the removal of the first ReLU layer is equivalent to $D_1^m \equiv {I_{d_1}}$, for which Theorem \ref{thm:impossible_reconstr} is then applicable.

%% file: tex/app/datasets.tex
\section{Details of Scenarios}\label{sec:app:datasets}
\input{tex/tables/scenarios.tex}
Table \ref{tab:scenarios} summarizes the general information of the datasets we cover in our experiments. In the following, we provide more details.

\noindent\textbf{Academic Benchmarks.} We choose the standard benchmark image datasets, i.e., CIFAR-10 \cite{Krizhevsky2009LearningML} and ImageNet \cite{Krizhevsky2012ImageNetCW}, which are considered in previous data reconstruction attacks. These two datasets originate from the machine learning community and are widely used as computer vision benchmarks for image classification and many other tasks. These two datasets mainly cover daily objects and show incremental complexity in various aspects (e.g., total pixels, color channels, class number).

\noindent\textbf{Medical Scenarios}. We consider three real-world medical imaging datasets made public by \cite{medmnist}, namely, RetinaMNIST, DermaMNIST, OrganMNIST. These three datasets corresponds to the tasks of intelligent diagnosis of iris-related, skin-related and organ-related pathology. We choose these three datasets out of the $9$ datasets from \cite{medmnist} based on its diversity in color channels and image variance. Besides, we use the ISIC skin cancer dataset \cite{Gutman2018SkinLA}, which consists of more high-resolution skin cancer images for evaluating our hybrid attack on deep CNNs.  

\noindent\textbf{Identity-Related Scenario.} We consider a face recognition system built with a subset of the Facescrub dataset \cite{Ng2014ADA}, which consists of portraits of $20$ celebrities randomly selected from the full dataset. 

%% file: tex/tables/scenarios.tex
\begin{table}[ht]
  \begin{center}
    \caption{Scenarios covered in experiments.}
    \label{tab:scenarios}
    \scalebox{0.7}{
    \begin{tabular}{llcc}
    \toprule
       Dataset & Task & Input Size ($d_0$) & \# Classes ($K$) \\
      \midrule
      CIFAR-10 \cite{Krizhevsky2009LearningML} & Object Classification & $3\times32\times32$ & $10$ \\ 
      FaceScrub \cite{Ng2014ADA} & Face Recognition & $3\times32\times32$ &  $20$ \\ 
    RetinaMNIST \cite{medmnist} & Iris Diagnosis & $3\times28\times28$ & $5$ \\
    DermaMNIST  \cite{medmnist}& Dermatology & $3\times28\times28$ & $7$ \\
    OrganMNIST \cite{medmnist} & Pathology & $1\times28\times28$ & $11$ \\
    \midrule
    ImageNet \cite{Krizhevsky2012ImageNetCW} & Object Classification & $3\times224\times224$ & $1000$ \\
    ISIC \cite{Gutman2018SkinLA} & Skin Cancer Diagnosis &  $3\times224\times224$ & $7$ \\

    \bottomrule
    \end{tabular}}
  \end{center}
\end{table}

%% file: tex/app/implementation.tex
\input{tex/alg.tex}

\section{More Details on Attack Implementations} \label{sec:app:config}
\noindent{\textbf{Dealing with the Bias Terms}.} It is easy to see, after the attacker has determined the $D_i^m$ and $\overline{g}_c^m$, the bias terms in the original gradient equation can be moved to the LHS, as a constant calibration to the ground-truth gradient. Specifically, w.l.o.g., considering the following single-sample gradient equation w.r.t. $W_i$ with bias terms, for $i = 1, \hdots, H$, we have
\begin{align}
    \overline{G}_i = \sum_{c}\overline{g}_c((D_{i}W_{i-1}...W_0X)([W_{H}]_c^{T}D_{H}...W_{i+1}D_{i+1}) \nonumber \\
    + (\sum_{j=0}^{i-1}(D_{i}W_{i-1}...W_{j+1}D_{j+1})b_j([W_{H}]_c^{T}D_{H}...W_{i+1}D_{i+1})))
\end{align}
where the constant term at the second line contains all the bias terms $(b_0, \hdots, b_{i-1})$ that have been multiplied with $W_i$ during the forwarding. By calibrating the ground-truth gradient $\overline{G}_i$ with its opposite, the gradient equation reduces to the form without bias term which we focus on in the main text. As final remarks, (i) the gradient equation w.r.t. $W_0$ has no bias term; (ii) the correctness of our algorithm for determining $\overline{g}_c^m$ and $D_i^m$ is independent from whether we consider the bias term or not, except for the process of determining $D_H^m$ from the gradient equation w.r.t. $b_{H-1}$.

\noindent{\textbf{Optimize the Attack Efficiency.}} The implementation of our attack algorithm on FCNs strictly follows our analysis, which makes the attack slightly more expensive than previous learning-based attacks, mainly because its additional cost in building and solving the large-scale sparse linear equation system. In our current implementations, we use multiprocessing to accelerate the building of the equation system, and use the sparse matrix representation to reduce the storage of the equations and leverage LSMR \cite{ChinLung2011LSMRAI} for fast equation-solving. How to further improve the efficiency of our theory-oriented attack may be a possible future direction.

\noindent{\textbf{Feature Matching vs. Gradient Matching.}} The feature-mapping problem in our hybrid attack is much cheaper and stabler compared with the gradient-matching problem in the following aspects: (i) Optimizing the gradient-matching problem requires the construction of the second-order computational graph to calculate the gradient of the model's gradient, which is likely to cause numeric instability \cite{Koh2017UnderstandingBP}, while the optimization of the feature-matching problem only requires to build the first-order computational graph for calculating the model's gradient, at a similar expense to a normal training procedure with better stability in computation; (ii) The optimization process on the gradient-matching problem requires the optimization itself to separate the signal for each single sample from the average gradient, while the feature-matching problem is much simplified as the feature maps reconstructed by our attack on FCN is already one-to-one correspondence to each data sample; (iii) Based on (ii), the optimization of the gradient-matching problem has to optimize the full batch of unknown samples (labels) as a whole, which has a huge demand on the GPU memory and the computing resources, while the optimization of the feature-matching problem only involves a single sample each time, which can be either done in parallel or in  sequence according to the available computing resources.

\noindent\textbf{Experiment Environments.} All the experiments are implemented with PyTorch \cite{NEURIPS2019_9015}, which is an open-source software framework for numeric computation and deep learning. All our experiments are conducted on a Linux server running Ubuntu 16.04, one AMD Ryzen Threadripper 2990WX 32-core processor and 1 NVIDIA GTX RTX2080 GPU.

%% file: tex/alg.tex
\section{Algorithm Details}
\label{sec:app:alg}
In this part, we provide the algorithmic descriptions of the key procedures in our proposed data reconstruction attacks on FCNs in Algorithm \ref{alg:determine_gcm}, \ref{alg:determine_label} \&  \ref{alg:determine_dm}.
\begin{algorithm}[ht]
\begin{algorithmic}[1]
\State{\textbf{Input:} The gradient of $W_H$, i.e., $\overline{G}_H$.}
\State{\textbf{Output:} Reconstructed labels $\{Y_1, \hdots, Y_M\}$ and loss vectors $\{(\overline{g}_c^{m})_{c=1}^{K}\}_{m=1}^{M}$.}
\State{Compute $r_c := [\overline{G}_H]_c/[\overline{G}_H]_1$ for every $c$ in $1, \hdots, K$.} 
\State{Find all the disjoint index groups $\{\mathcal{I}^m\}_{m=1}^{M}$ where $(r_2)_j$ is constant whenever $j\in\mathcal{I}^m$.}
\Comment{$M$ is hence the inferred batch size and $\mathcal{I}^m$ is the index set of the exclusively activated neurons at the last ReLU layer.}

\ForAll{$c$ in $1, \hdots, K$}
\ForAll{$m$ in $1, \hdots, M$}
\State{Select an arbitrary index $j$ from $\mathcal{I}^m$.}
\State{$\overline{g}_c^{m}/\overline{g}_1^{m} \gets [r_c]_j$.}
\EndFor
\EndFor

\ForAll{$m$ in $1, \hdots, M$}
\State{$Y_m \gets$ Apply Algorithm \ref{alg:determine_label} to $(\overline{g}_c^{m})_{c=1}^{K}$}.
\State{Estimate the upper bound of feasible range of $\overline{g}_1^{m}$ as $\delta_m \gets \overline{g}_{1}^{m}/\overline{g}_{Y_m}^{m}$}
\State{Fix $\overline{g}_{1}^{m} = 2\times{\delta_m}/3$.} \Comment{This is practiced in all our experiments.}
\State{Calculate each $\overline{g}_c^{m}$ according to the ratio.}
\EndFor

 \caption{Determine $\{(\overline{g}_c^{m})_{c=1}^{K}\}_{m=1}^{M}$.}
 \label{alg:determine_gcm}
 \end{algorithmic}
\end{algorithm}

\begin{algorithm}[ht]
\begin{algorithmic}[1]
\State{\textbf{Input:} The loss vector for the $m$-th sample $(\overline{g}_c^{m})_{c=1}^{K}$.}
\State{\textbf{Output:} Reconstructed label $Y_m$.}
\If{$(\overline{g}_c^{m})_{c=1}^{K}$ have one negative element} \\
\Return{$Y_m \gets$ The index of the negative element}
\Else \\ 
{\Return{$Y_m \gets 1$}}
\EndIf 
\caption{Exact label reconstruction from the loss vector.}
 \label{alg:determine_label}
 \end{algorithmic}
\end{algorithm}

\begin{algorithm}[ht]
\begin{algorithmic}[1]
\State{\textbf{Input:} The gradients $(\overline{G}_i)_{i=0}^{H}$ at each layer, the index sets $(\mathcal{I}^m_{H})_{m=1}^{M}$ of exclusively activated neurons at the last ReLU layer and the reconstructed $\{(\overline{g}_c^m)_{c=1}^{K}\}_{m=1}^{M}$}
\State{\textbf{Output:} Reconstructed activation patterns. $\{(D_i^{m})_{i=1}^{H}\}_{m=1}^{M}$.}
\State{$\mathcal{I}_{\text{cur}} \gets \{\mathcal{I}^m_{H}\}_{m=1}^{M}$.}
\ForAll{$i$ in $H-1, \hdots, 1$}
\ForAll{$m$ in $1, \hdots, M$}
\State{Select an arbitrary index $j$ from $\mathcal{I}^m_{\text{cur}}$.}
\State{$\text{diag}(D_i^{m}) \gets ([\overline{G}_i]_{:,j} \neq 0)$}
\EndFor
\State{Construct the index sets $\{\mathcal{I}^m_{i}\}_{m=1}^{M}$ of exclusively activated neurons at the $i$-th layer from $\{D_i^{m}\}_{m=1}^{M}$.}
\State{$\mathcal{I}_{\text{cur}}\gets\{\mathcal{I}^m_{i}\}_{m=1}^{M}$.}
\EndFor

\State{Solve $D_H^{m}$ from the binary equation $\frac{1}{M}\sum_{m=1}^M\sum_{c=1}^{K}\overline{g}_c^m[W_H]_c^TD_H^mI_{d_H} = \frac{\partial{\ell}}{\partial{b_{H-1}}}$.}
\caption{Determine activation patterns $\{(D_i^{m})_{i=1}^{H}\}_{m=1}^{M}$.}

 \label{alg:determine_dm}
 \end{algorithmic}
\end{algorithm}

%% file: tex/app/eval.tex
\section{More Evaluation Results}
\label{sec:app:exp}

\begin{figure}[ht]
\begin{center}
\includegraphics[width=0.4\textwidth]{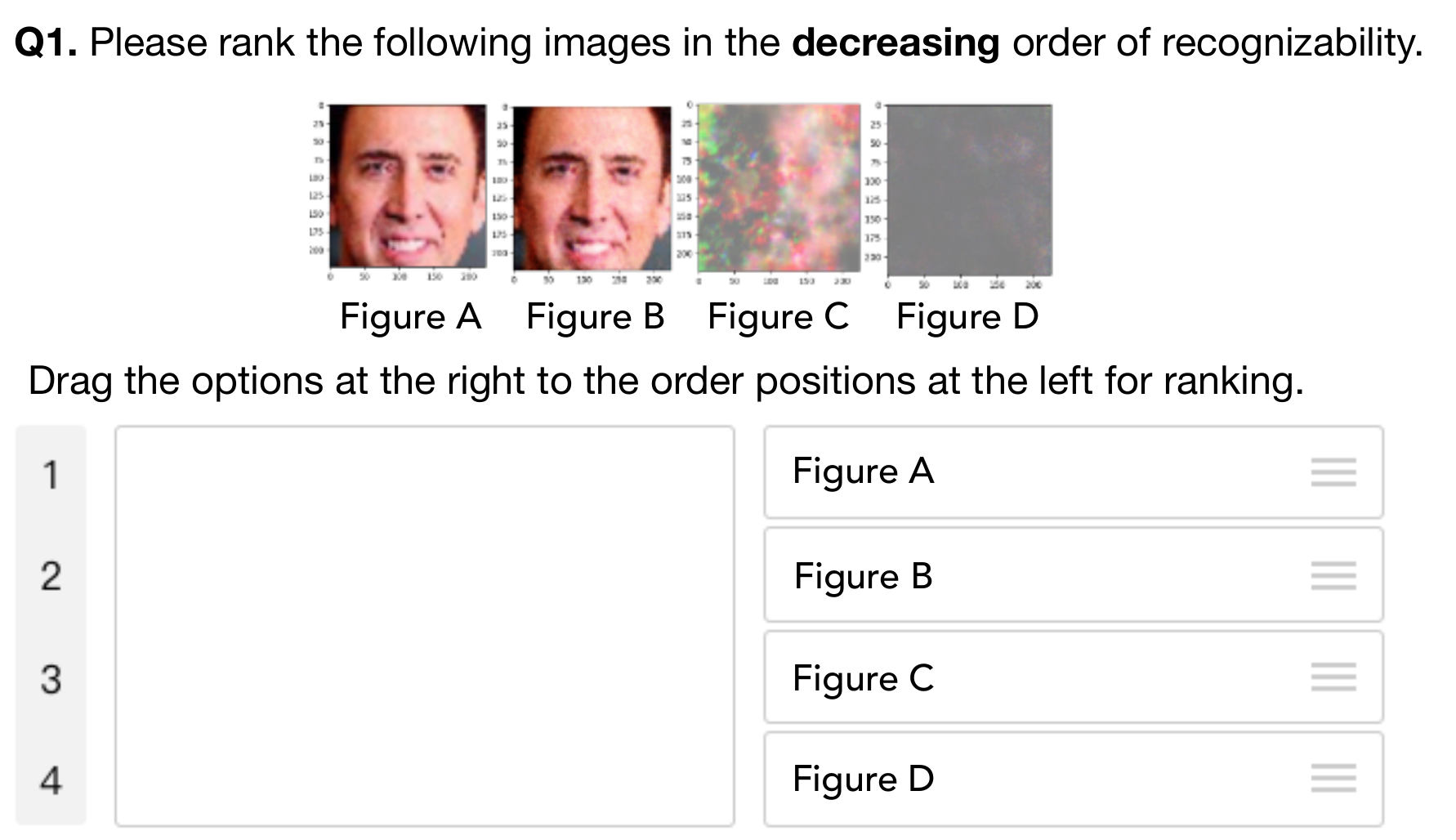}
\caption{A sample question from our survey for human evaluation of reconstruction quality.}
\label{fig:ac8_sample_question}
\end{center}
\end{figure}

\begin{figure}[ht]
\begin{center}
\includegraphics[width=0.4\textwidth]{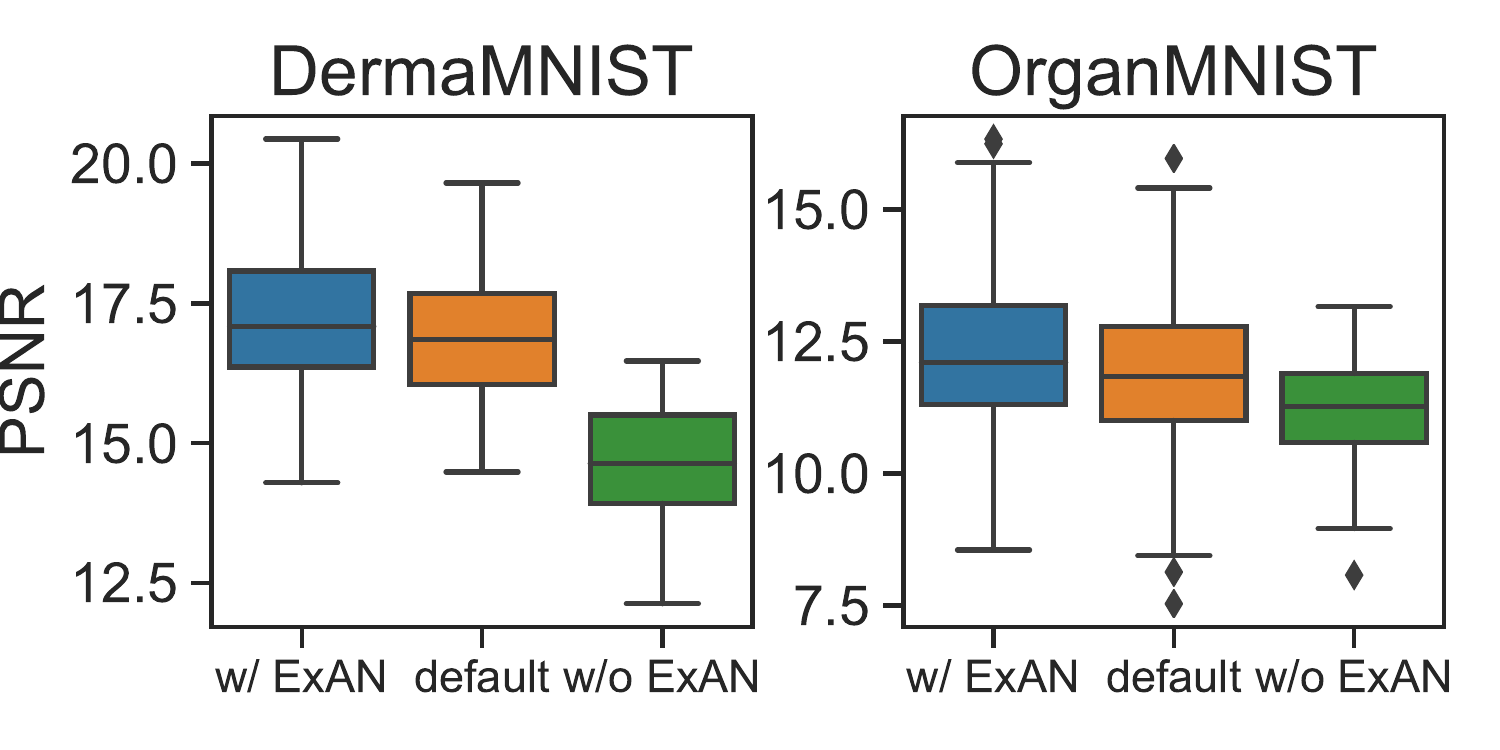}
\caption{Omitted results on other datasets for Fig. \ref{fig:b4_exan_asr}.}
\label{fig:app:b4_exan_asr}
\end{center}
\end{figure}

\begin{figure*}
\begin{center}
\includegraphics[width=0.7\textwidth]{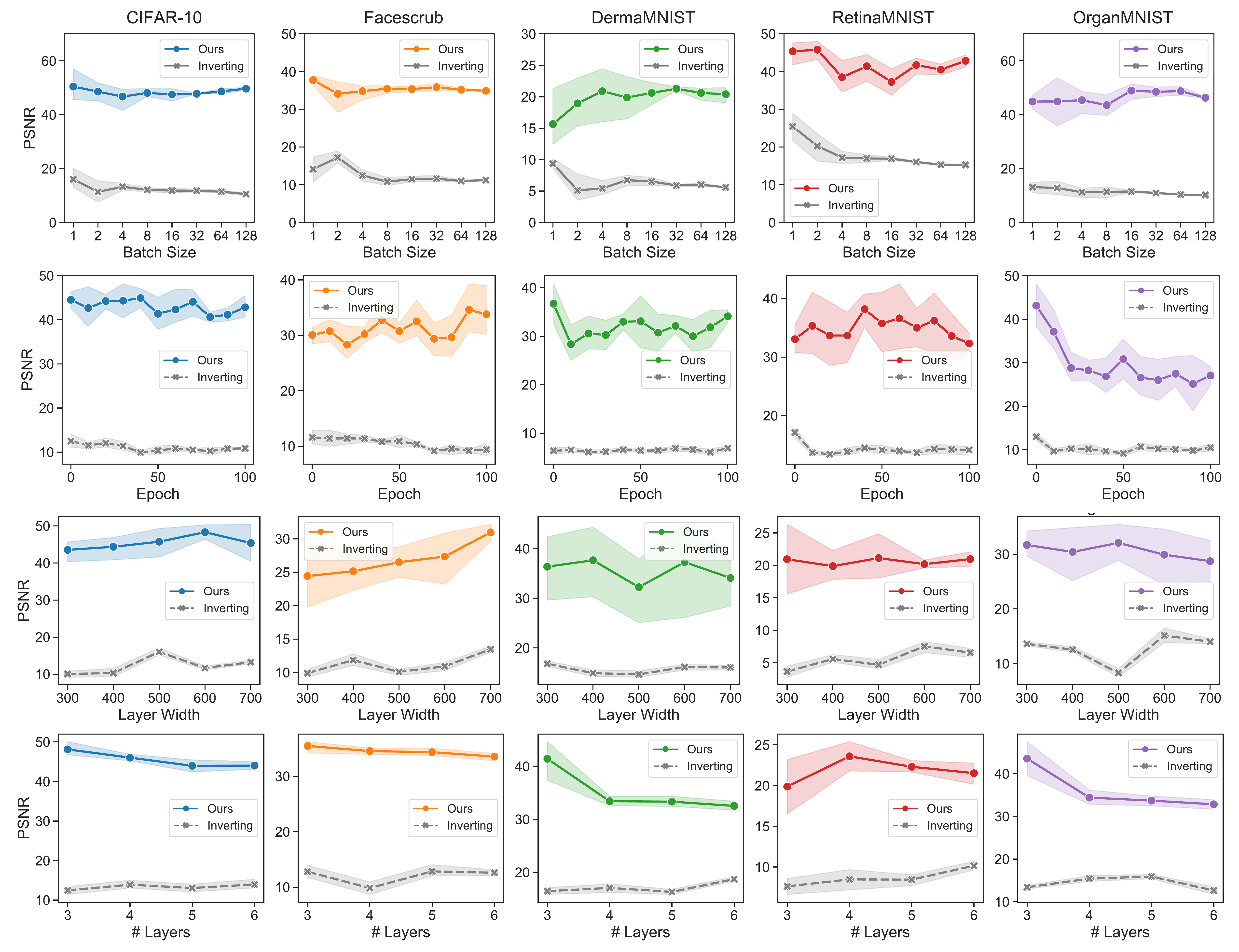}
\caption{Omitted results on other datasets for Fig. \ref{fig:abl_suite}.}
\label{fig:app:abl_suite}
\end{center}
\end{figure*}

\begin{figure*}
\begin{center}
\includegraphics[width=1.0\textwidth]{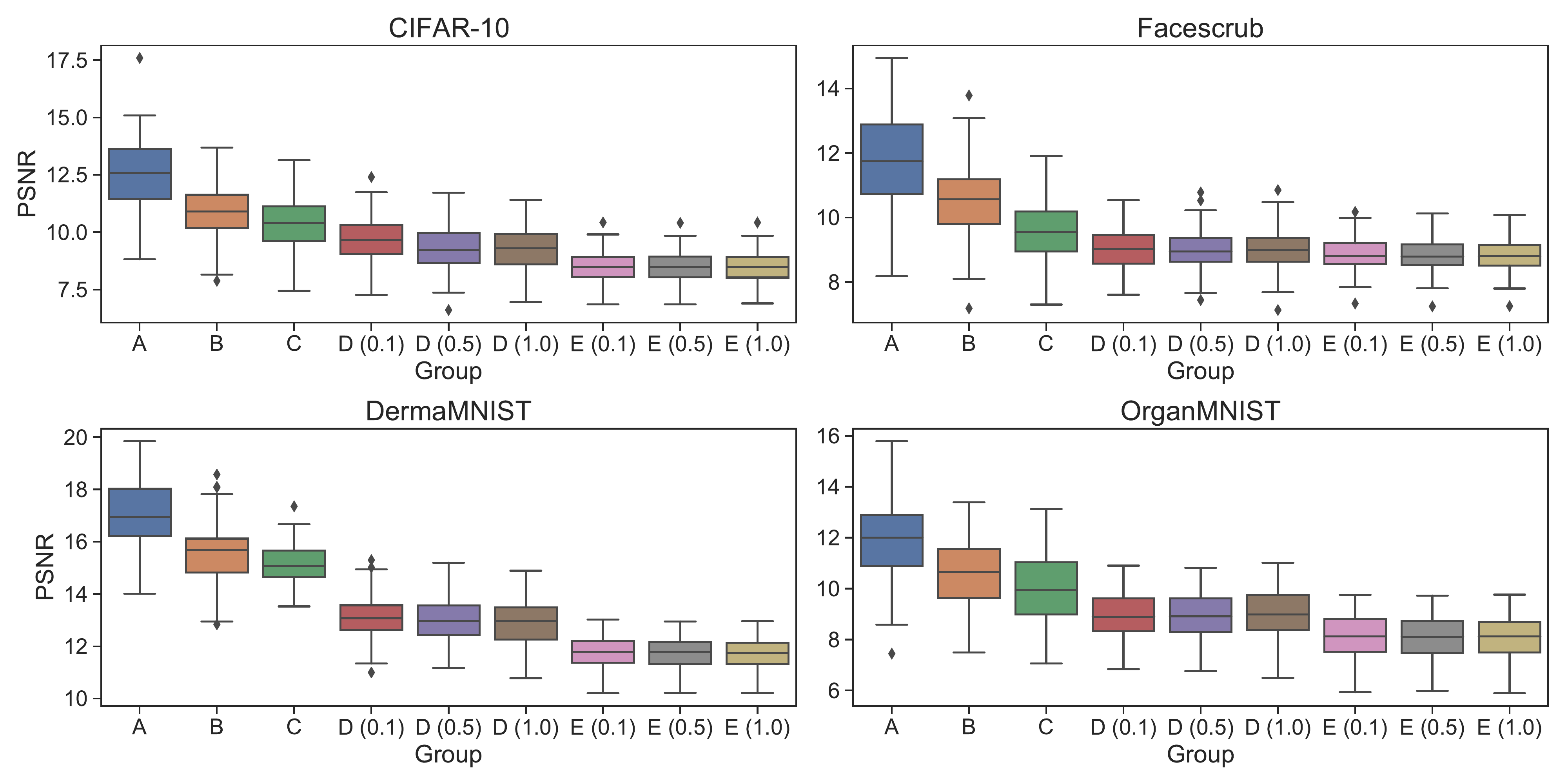}
\caption{Omitted results on other datasets for Fig. \ref{fig:ac5_defense}.}
\label{fig:app:ac5_defense}
\end{center}
\end{figure*}

\noindent\textbf{Omitted Results on Other Datasets.} In Fig. \ref{fig:app:abl_suite}, we present the omitted results accompanying Fig. \ref{fig:abl_suite} in the main text. In Fig. \ref{fig:app:ac5_defense}, we present the omitted results accompanying Fig. \ref{fig:ac5_defense} in the main text. In Fig. \ref{fig:app:b4_exan_asr}, we present the omitted results accompanying Fig. \ref{fig:b4_exan_asr} in the main text.

\begin{figure}[ht]
\begin{center}
\includegraphics[width=0.5\textwidth]{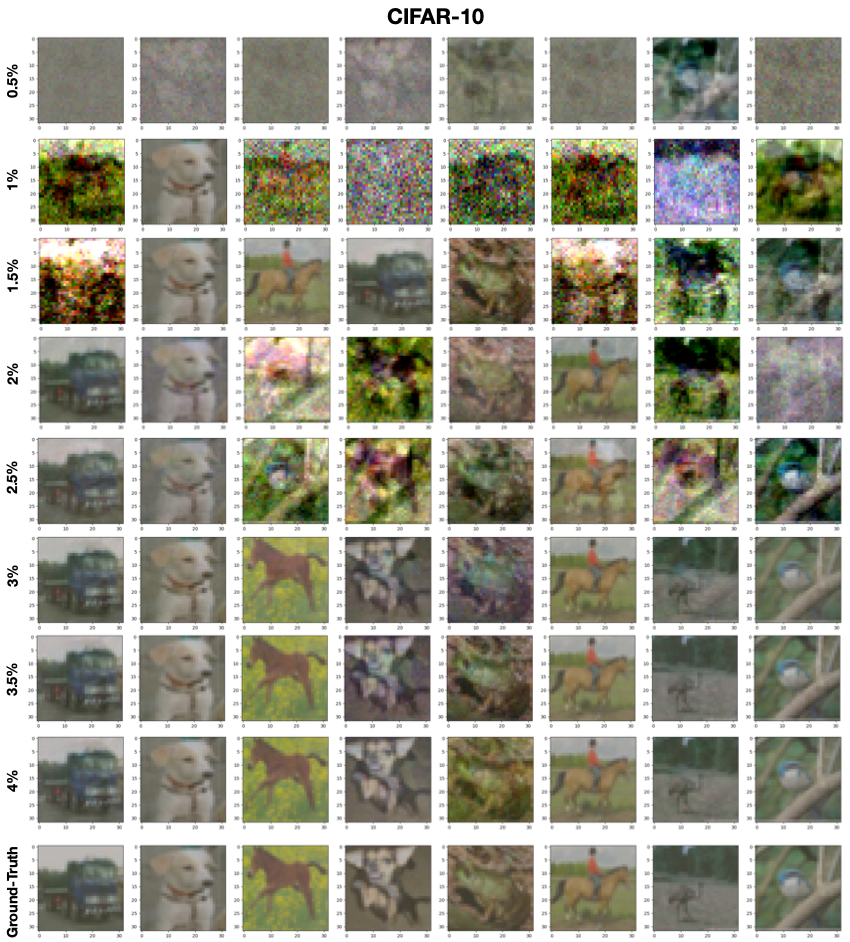}
\caption{Reconstruction results of our proposed attack algorithm when different proportions of gradient information are provided.}
\label{fig:scale_demo}
\end{center}
\end{figure}

\begin{figure}[ht]
\begin{center}
\includegraphics[width=0.45\textwidth]{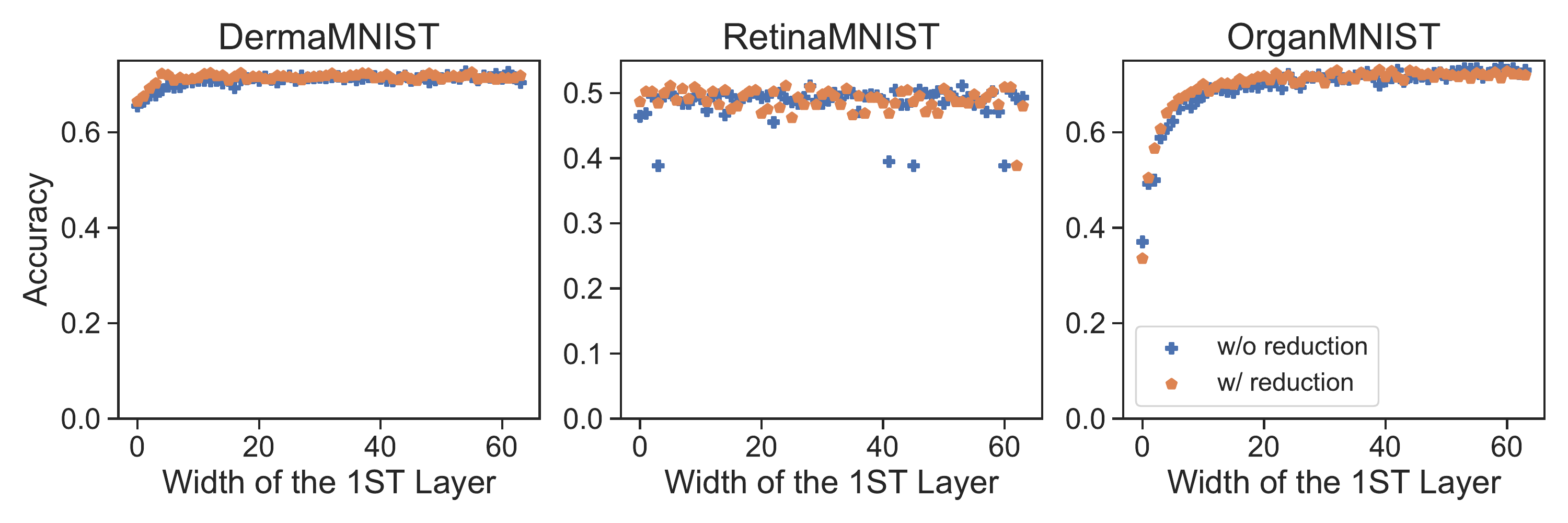}
\caption{The accuracy of three-layer FCNs on the main task of DermaMNIST, RetinaMNIST and OrganMNIST, with and without exclusivity reduction. The width of the first ReLU layer varies in $[1, 64]$.}
\label{fig:acc_plot}
\end{center}
\end{figure}

\begin{figure*}
\begin{center}
\includegraphics[width=1.0\textwidth]{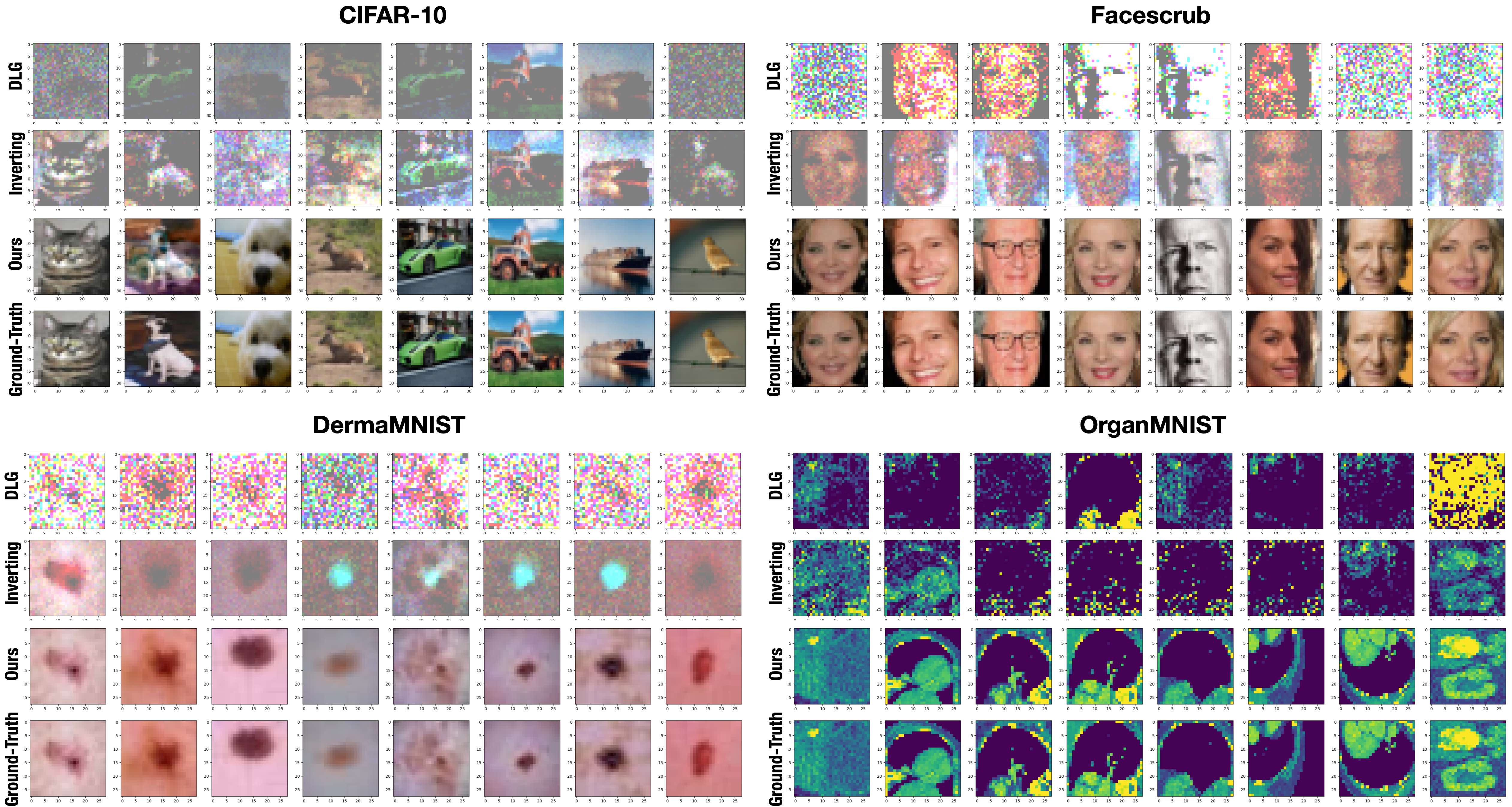}
\caption{Omitted visualization results accompanying Fig. \ref{fig:demo_body} in the main text.}
\label{fig:app:demo_body}
\end{center}
\end{figure*}

\begin{figure*}
\begin{center}
\includegraphics[width=1.0\textwidth]{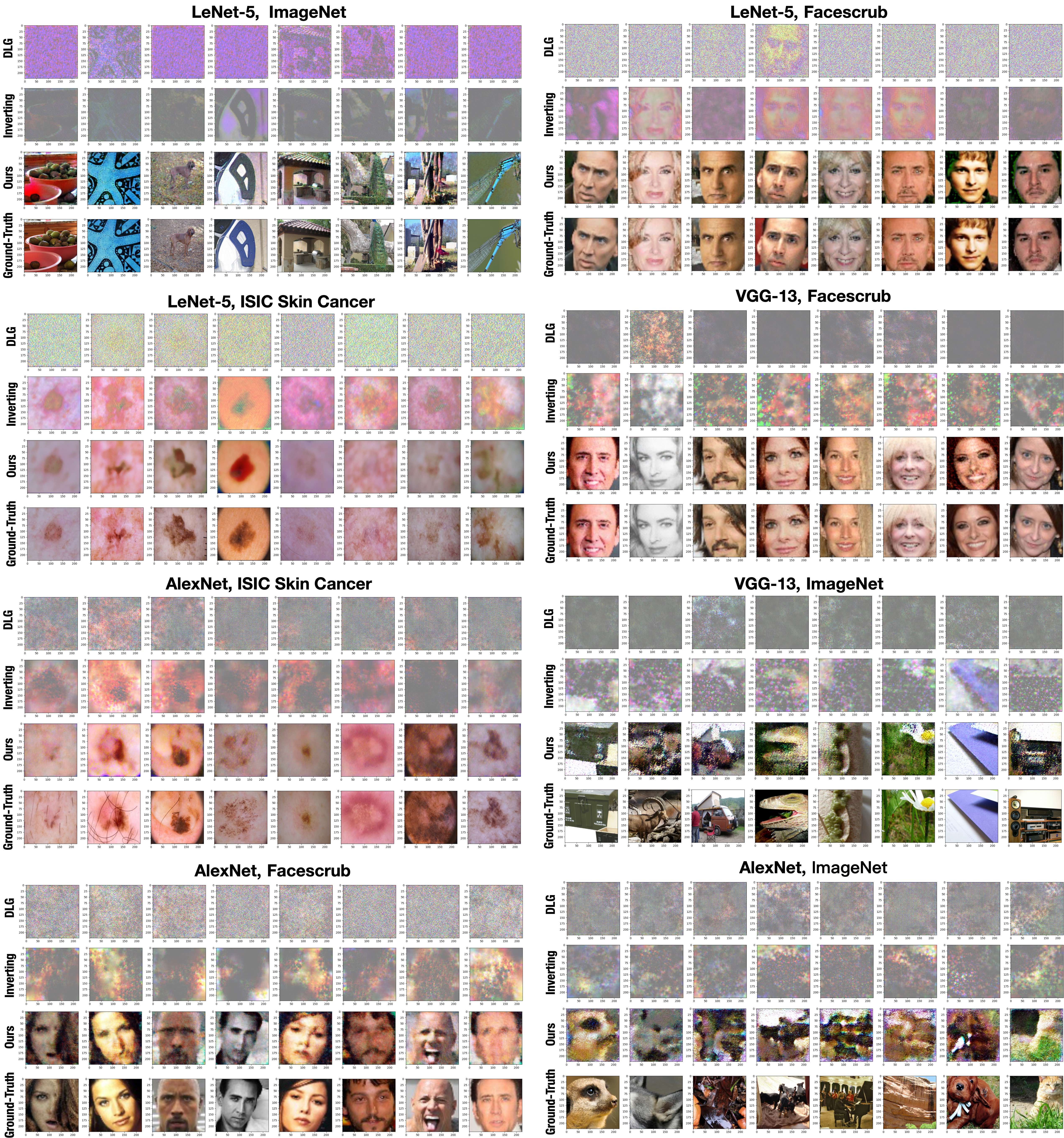}
\caption{Omitted visualization results accompanying Fig. \ref{fig:ac2_vgg_demo} in the main text.}
\label{fig:app:dcnn_results}
\end{center}
\end{figure*}

\noindent\textbf{Omitted Visualization Results.} In Fig. \ref{fig:app:demo_body}, we present the omitted results accompanying Fig. \ref{fig:demo_body} in the main text. In Fig. \ref{fig:app:dcnn_results}, we present the omitted results accompanying Fig. \ref{fig:ac2_vgg_demo} in the main text.

\noindent\textbf{More Details on Human Evaluation.} First, we collect a group of reconstruction results of DLG, Inverting and our attack on the same batch for each of the following $8$ test cases, namely, FCN on Facescrub \& CIFAR-10, LeNet-5 on Facescrub \& CIFAR-10, AlexNet on ImageNet \& Facescrub, and VGG-13 on ImageNet \& Facescrub, where the batch size is always set as $8$. We do not include the reconstruction results on ISIC skin cancer dataset because the images may be inappropriate for all of our participants to view. With the collected reconstruction results, we prepare a survey composed of $24$ questions in the same format. As shown in Fig. \ref{fig:ac8_sample_question}, each question shows $4$ images in a line (i.e., $3$ reconstruction results for the same ground-truth image and the corresponding ground-truth image, positioned in a random order). For each question, the participants are required to rank the $4$ images in a decreasing order of recognizability.

\begin{figure*}
\begin{center}
\includegraphics[width=1.0\textwidth]{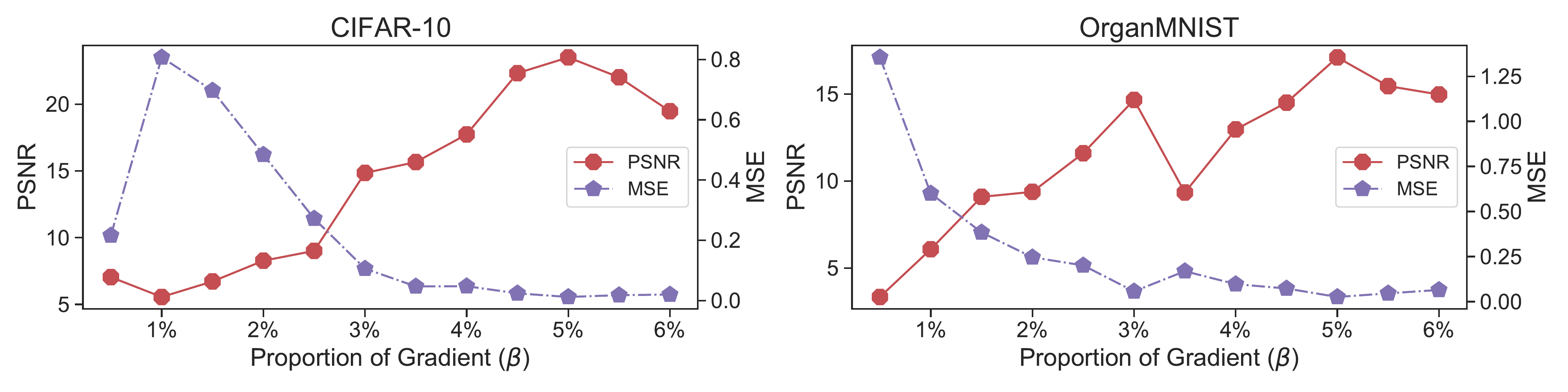}
\caption{Impact of the proportion of gradients accessed by the attacker on the reconstruction accuracy.}
\label{fig:scale}
\end{center}
\end{figure*}

\noindent{\textbf{Scale of Leaked Gradient Information.}} In this part, we vary the proportion of the ground-truth gradients accessed by the adversary to analyze how the level of gradient information leakage impacts the reconstruction results. We set the target model as a $3$-layer FCN ($d_0$-$512$-$K$). Specifically, we randomly choose $\beta$ proportion of weight parameters. After the loss vectors and the activation patterns are determined, we only allow the attack algorithm to use this part of gradients to form the linear gradient equation system. Fig. \ref{fig:scale} plots the changes of PSNR and MSE of our proposed attack algorithm when $\beta$ increases. As we can see from Fig. \ref{fig:scale}, with an increased proportion of gradient information accessed by the adversary, the performance of data reconstruction attacks is strengthened correspondingly. For example, when $\beta$ increases from $0.5\%$ to $6\%$, the PSNR on CIFAR-10 increases from $7.04$ to over $20$. Fig. \ref{fig:scale_demo} in \ref{sec:app:exp} provides the visualization results. This confirms our statement in Theorem \ref{col:optim_bound_mult}: Once the sufficient exclusivity condition is satisfied, the MSE of data reconstruction attacks would decrease quadratically.

\begin{figure*}[ht]
\begin{center}
\includegraphics[width=1.0\textwidth]{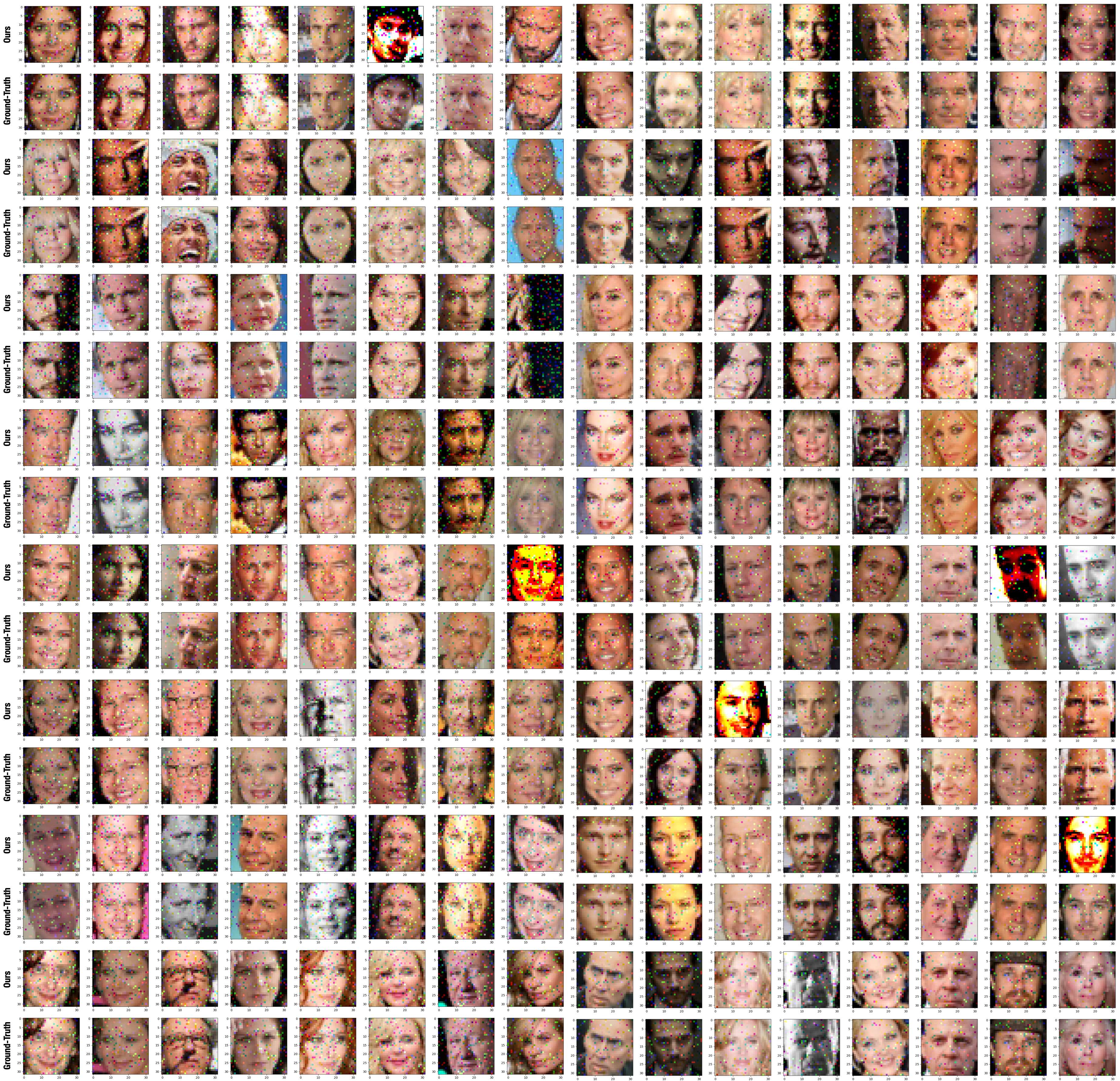}
\caption{Visualization of a reconstructed batch of $128$ samples from Facescrub, where the ground-truth images are slightly perturbed to satisfy the condition of sufficient exclusivity.}
\label{fig:ac1_size_demo_facescrub}
\end{center}
\end{figure*}

